\documentstyle[12pt,epsfig,a4]{article} 
\newcommand{\vs}{\vspace{-0.25cm}}
\begin{document}
\begin{center}
{\Large{\bf Single-particle potential from resummed  ladder 
diagrams}\footnote{Work supported in part by DFG and NSFC (CRC 110).}}
\bigskip

N. Kaiser\\
\medskip
{\small Physik-Department T39, Technische Universit\"{a}t M\"{u}nchen,
    D-85747 Garching, Germany

\smallskip
{\it email: nkaiser@ph.tum.de}}
\end{center}
\medskip
\begin{abstract}
A recent work on the resummation of fermionic in-medium ladder diagrams to all 
orders is extended by calculating the complex single-particle potential 
$U(p,k_f)+ i\,W(p,k_f)$ for momenta $p<k_f$ as well as $p>k_f$. The on-shell
single-particle potential is constructed by means of a complex-valued 
in-medium loop that includes corrections from a test-particle of momentum $\vec p$
added to the filled Fermi sea. The single-particle potential $U(k_f,k_f)$ at the 
Fermi surface as obtained from the resummation of the combined 
particle and hole ladder diagrams is shown to satisfy the Hugenholtz-Van-Hove 
theorem. The perturbative contributions at various orders $a^n$ in the scattering 
length are deduced and checked against the known analytical results at order $a^1$ 
and $a^2$. The limit $a\to\infty$ is studied as a special case and a 
strong momentum dependence of the real (and imaginary) single-particle potential 
is found. This indicates an instability against a phase transition to a state with an 
empty shell inside the Fermi sphere such that the density gets reduced by about 
$5\%$. For comparison, the same analysis is performed for the resummed 
particle-particle ladder diagrams alone. In this truncation an instability for 
hole-excitations near the Fermi surface is found at strong coupling. For the 
set of particle-hole ring diagrams the single-particle potential is calculated as 
well. Furthermore, the resummation of in-medium ladder diagrams to all orders is 
studied for a two-dimensional Fermi gas with a short-range two-body contact-interaction.  
\end{abstract}

%\bigskip

PACS: 05.30.Fk, 12.20.Ds, 21.65+f, 25.10.Cn
%\vspace{-0.5cm}
\section{Introduction and summary}
Dilute degenerate many-fermion systems with large scattering lengths $a$ are of 
interest, e.g. for modeling the low-density behavior of nuclear and neutron 
star matter. Because of the possibility to tune magnetically atomic interactions, 
ultracold fermionic gases provide an exceptionally valuable tool to explore the 
non-perturbative many-body dynamics (for a recent comprehensive 
review of this field, see ref.\cite{zwerger}). Of particular interest 
in this context is the so-called unitary limit, in which the two-body interaction 
has the critical strength to support a bound-state at zero energy. As a 
consequence of the diverging scattering length, $a\to \infty$, the strongly 
interacting many-fermion system becomes scale-invariant. Its ground-state energy 
is determined by a single universal number, the so-called Bertsch parameter 
$\xi$, which measures the ratio of the energy per particle $\bar E(k_f)^{(\infty)}$ 
to the (free) Fermi gas energy, $\bar E(k_f)^{(0)}=3k_f^2/10M$. Here, $k_f$ denotes 
the Fermi  momentum and $M$ the large fermion mass.

In a recent work \cite{resum1} the complete resummation of the combined 
particle-particle and hole-hole ladder diagrams generated by a 
contact-interaction proportional to the scattering length $a$ has been achieved. 
A key to the solution of this (restricted) problem has been a different 
organization of the many-body calculation from the start. Instead of treating 
(propagating) particles and holes separately, these are kept together and the 
difference to the propagation in vacuum is measured by a ``medium-insertion''. This 
approach is founded in the following identical rewriting of the (non-relativistic) 
particle-hole propagator:
\begin{eqnarray} {i\, \theta(|\vec p_j|-k_f) \over \omega_j -\vec p_j^{\,2}/2M+ 
i\epsilon} + {i\, \theta(k_f-|\vec p_j|) \over \omega_j -\vec p_j^{\,2}/2M- 
i\epsilon} = {i \over \omega_j -\vec p_j^{\,2}/2M+ i\epsilon} -2\pi \, 
\delta(\omega_j -\vec p_j^{\,2}/2M)\, \theta(k_f-|\vec p_j|)\,, \end{eqnarray}
for an internal fermion-line carrying energy $\omega_j$ and momentum $\vec p_j$. In that 
organizational scheme the pertinent in-medium loop  is complex-valued, 
and therefore the contribution to the energy per particle $\bar E(k_f)$ at order 
$a^n$ is not directly obtained from the $(n-1)$-th power of the in-medium loop. 
However, after reinstalling the symmetry factors $1/(j+1)$  which belong to 
diagrams with $j+1$ double medium-insertions, a real-valued expression is 
obtained for all orders $a^n$. Known results about the low-density expansion 
\cite{hammer,steele} up to and including order $a^4$ could be reproduced with 
improved numerical accuracy. The emerging series in $a k_f$ could even be summed 
to all orders in the form of a double-integral over an arctangent-function. In 
that explicit representation the limit $a\to \infty$ could be taken 
straightforwardly and the value $\xi_n=0.507$ was found for the ``normal'' 
Bertsch parameter \cite{resum1}. This number is to be compared with the value 
$\xi_n^{(pp)}\simeq 0.237$ resulting from a resummation of the particle-particle 
ladder diagrams only \cite{schaefer}. Interestingly, extrapolations of 
experimentally measured thermodynamical quantities of a gas of $^6$Li-atoms at 
a Feshbach resonance from finite to zero temperature, which smoothly pass over 
the pairing transition at $T_c\simeq T_f/6 $, give indications for a Bertsch 
parameter $\xi_n \simeq 0.45$ in the normal phase  (see Fig.\,3 in ref.\cite{zwierlein}). 
The true behavior close to zero temperature is governed by pairing, which leads to 
$\xi = 0.376$. In comparison to the extrapolated value $\xi_n$, the finding
$\xi_n=0.507$ of ref.\cite{resum1} appears to be quite good. The resummation 
of fermionic in-medium ladder diagrams to all orders has been extended in 
ref.\cite{resum2} by considering the effective range in the s-wave interaction and 
a (spin-independent) p-wave contact interaction. The generalization to a binary 
many-fermion system with two different scattering lengths, $a_s$ and $a_t$, has 
also been treated.

The purpose of the present paper is to further advance the non-perturbative 
resummation by calculating in this framework the complex single-particle 
potential. With its separate dependence on momentum $p$ and density $\rho=
k_f^3/3\pi^2$, the on-shell single-particle potential $U(p,k_f)+ i\,
W(p,k_f)$ provides more detailed information about the involved many-body 
dynamics than the interaction energy per particle $\bar E(k_f)$. In particular, 
the single-particle spectrum allows one to reveal inherent instabilities. 
Generally, the single-particle potential can be obtained by computing the first 
functional derivative of the energy density with respect to the occupation density. 
In practice such a functional derivative is performed by adding a test-particle of momentum $\vec p$ 
to the filled Fermi sea. The momentum regions $p<k_f$ and $p>k_f$, 
corresponding to hole and particle excitations, require yet separate a treatment.

The present paper is organized as follows: In section 2, the modifications of 
the complex-valued in-medium loop which arise from the introduction of a 
test-particle with momentum $\vec p$ are calculated first. The resulting 
functions (depending on three dimensionless variables) provide the essential 
technical tool to construct the complex-valued single-particle potential 
in an efficient way from the expression for the resummed interaction density. 
One obtains concise double-integral representations for $U(p,k_f)$ and $W(p,k_f)$ as 
derived from the complete resummation of the combined particle-particle and hole-hole 
ladder diagrams. The different integration domains involved for $p<k_f$ and 
$p>k_f$ arrange automatically for the correct continuation across the Fermi surface 
$p=k_f$, which is continuous but not smooth. The validity of the 
Hugenholtz-Van-Hove theorem, which relates the single-particle potential 
$U(k_f,k_f)$ at the Fermi surface $p=k_f$ to the interaction energy per particle 
$\bar E(k_f)$, is verified explicitly for the present non-perturbative calculation. 
Rather subtle $\delta$-function terms, which arise from differentiating the 
discontinuity of the arctangent-function at infinity, play a crucial role for 
the actual numerical validity of the Hugenholtz-Van-Hove theorem. In the next step 
the perturbative contributions to $U(p,k_f)$ and  $W(p,k_f)$ at several orders 
$a^n$ in the scattering length are deduced and checked against the known analytical 
results at order $a^1$ and $a^2$. Results for the momentum dependence are presented 
up to 5-th order. The limit $a\to \infty$ is studied as a special case and 
a strong  momentum dependence of the potential $U(p,k_f)^{(\infty)}$ is found. 
This feature indicates an instability against a phase transition to a state with an empty 
shell inside the Fermi sphere (a so-called Sarma phase \cite{sarma}) such that the 
density gets reduced by about $5\%$.

Section 3 is devoted to the same analysis of the particle-particle ladder 
diagrams which can be resummed to all orders in the form of a geometrical 
series. Results for the momentum dependence of the real and imaginary 
single-particle potential are presented up to 5-th order in $a$. In the case of 
the truncated particle-particle ladder series an instability for the excitation 
of holes states with momenta $3k_f/4 < p <k_f$ is found in the limit 
$a\to \infty$. The continuation of the complex-valued single-particle 
potential $U(p,k_f)+i\,W(p,k_f)$ into the region outside the Fermi surface $p>k_f$ 
requires in this case several modifications of the formalism. In section 4 the 
real single-particle potential $U(p,k_f)$ arising from particle-hole ring diagrams 
is calculated and results are presented from third up to 8-th order in $a$.   

In the appendix, the in-medium ladder diagrams for a two-dimensional Fermi gas 
with a short-range two-body contact-interaction are studied and the corresponding 
resummed interaction energy per particle $\bar E(k_f)$ is calculated.  

It should be stated clearly that a realistic description of the unitary Fermi gas is 
neither possible nor intended in the present framework. For the unitary Fermi gas at 
zero temperature pairing plays a major role and the pairing field and the ordinary 
mean-field are of comparable magnitude \cite{mag}. The present results should be viewed 
as model studies on the way to a more complete treatment. 

\section{Complex single-particle potential from  
ladder diagrams}  
In perturbation theory the energy density of an interacting many-fermion system 
is represented by closed multi-loop diagrams with medium-insertions \cite{resum1} 
accounting for the presence of the filled Fermi sea. By computing the first functional 
derivative of the energy density with respect to the occupation density 
one obtains the single-particle self-energy generated by the interactions with the 
homogeneous fermionic medium. At a practical level this method amounts 
to adding a test-particle\footnote{The computation of 
the first functional derivative by adding an infinitesimal test-particle 
represents a constructive approach to the on-shell single-particle potential. The 
agreement with the conventional particle-hole counting scheme has been checked for 
model interactions at first and second order in ref.\cite{jeremy}.}  with momentum 
$\vec p$ to the filled Fermi sea, via the substitution \cite{pionpot}: 
\begin{equation} \theta(k_f-|\vec p_j|)\,\,\to\,\, N_\eta(\vec p_j,\vec p\,) =
\theta(k_f-|\vec p_j|)+ 4\pi^3 \eta \,\delta^3(\vec p_j-\vec p\,)\,,\end{equation} 
with $\eta$ an infinitesimal parameter. For creating a hole-excitation (with 
momentum $p<k_f$) or a particle-excitation (with momentum $p>k_f$) one sets 
$\eta =\mp 1/V$, with $V$ the large volume of the system. To linear order in $\eta$ 
the interaction energy density changed then as:
\begin{equation}{k_f^3 \over 3\pi^2}\bar E(k_f) \,\,\to \,\, {k_f^3 \over 3\pi^2}
\bar E(k_f) + \eta \Big[ U(p,k_f)+i\, W(p,k_f)\Big] \,, \end{equation} 
where $\bar E(k_f)$ is the interaction energy per particle, and 
$U(p,k_f)+i\,W(p,k_f)$ denotes the complex-valued single-particle potential. Its 
imaginary part determines the decay width $\Gamma$ of a hole or particle excitation 
by $\Gamma = 2\,{\rm sign}(k_f-p)\, W(p,k_f)$. In the following this construction of 
the on-shell single-particle potential $U(p,k_f)+i\,W(p,k_f)$ is carried out for 
the combined particle-particle and hole-hole ladder diagrams \cite{resum1} which 
are generated by a contact-interaction proportional to the s-wave scattering length 
$a$ to all orders. The sign-convention for $a$ is chosen is such that a positive 
scattering length $a>0$ corresponds to attraction. In order to evade the pairing 
instability in many-fermion systems with attractive interactions, all following 
non-perturbative expressions should be applied to repulsive scattering lengths $a<0$
only.  
\subsection{Modifications of the in-medium loop}
\begin{figure}
\begin{center}
\includegraphics[scale=0.5,clip]{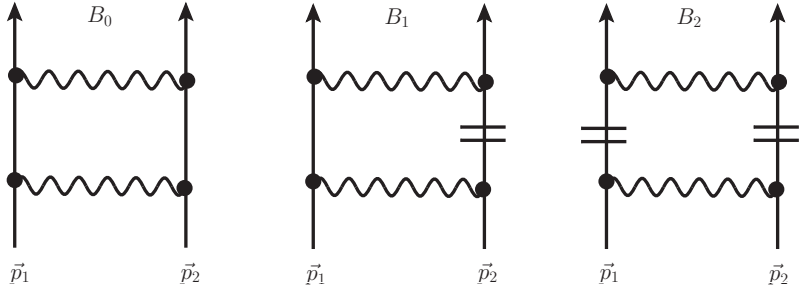}
\end{center}
\vspace{-.5cm}
\caption{On-shell in-medium loop organized in the number of medium-insertions 
({\bf =}). The middle diagram has a reflected partner. The external momenta 
$\vec p_1$ and $\vec p_2$ are unconstrained.}
\end{figure}

The basic quantity in order to achieve the resummation of ladder diagrams to 
all orders in ref.\cite{resum1} has been the complex-valued in-medium loop. 
Therefore, we calculate first the modifications of the in-medium loop 
which arise from the introduction of the test-particle with momentum $\vec p$.  
It is convenient to work with the half-sum $\vec P = (\vec p_1+\vec p_2)/2$ and 
half-difference $\vec q = (\vec p_1-\vec p_2)/2$ of two external momenta $\vec p_1$ 
and $\vec p_2$. In contrast to the calculation of the interaction energy per particle 
$\bar E(k_f)$ in ref.\cite{resum1}, the external momenta $\vec p_1$ and $\vec p_2$ 
are now not constrained to lie inside a Fermi sphere of radius $k_f$. The real part 
of the on-shell in-medium loop comes from the (middle) diagram in Fig.\,1 with one 
medium-insertion and after the substitution specified in eq.(2) it reads:
\begin{equation} {\rm Re}\,B_1= -4\pi a \int\!\!{d^3 l \over (2\pi)^3}
\, {1\over \vec l^{\,2}- \vec q^{\,2}} \Big\{N_\eta(\vec P+\vec l, \vec p\,)+ 
N_\eta(\vec P-\vec l, \vec p\,)\Big\} \,.\end{equation} 
The contribution linear in $\eta$ is simply twice the same energy denominator 
and after averaging over the directions of $\vec p$ one finds:\footnote{Since 
only terms linear in $\eta$ are relevant, this averaging can be done at an early 
stage of the calculation.} 
\begin{equation}{\rm Re}\,\bar B_1= -{a k_f \over \pi} \Big\{ R(s,\kappa) + 
\tilde \eta \, \widehat R(s,\kappa,x)\Big\}\,, \end{equation} 
with the dimensionless infinitesimal parameter $\tilde \eta= \eta \,\pi^2/k_f^3$ and 
the logarithmic functions:
\begin{equation} R(s,\kappa) = 2 +{1\over 2s}[1-(s+\kappa)^2]\ln{1+s +\kappa
\over |1-s -\kappa|}+{1\over 2s}[1-(s-\kappa)^2]\ln{|1+s -\kappa|
\over |1-s +\kappa|}\,, \end{equation} 
\begin{equation} \widehat R(s,\kappa,x) ={1\over s x} \ln{|(s+x)^2-\kappa^2|\over 
|(s-x)^2-\kappa^2|} \,,\qquad \qquad \widehat R(s,\kappa,0) ={4\over 
s^2-\kappa^2} \,, \end{equation}
written in terms of the dimensionless variables $s = P/k_f$, $\kappa =  q/k_f$ and 
$x = p /k_f$. Actually, the function $R(s,\kappa)$ will be needed later only in 
regions where $ 1+s -\kappa$ and $1-s +\kappa$ are both positive. Therefore, one 
could drop the absolute magnitudes in the argument of the second logarithm in 
eq.(6). An interesting relation between these functions is: $\int_0^1 dx\,x^2 
\widehat R(s,\kappa,x) =  R(s,\kappa)$.

Next, we consider the imaginary part of the on-shell in-medium loop. It is generated 
by all three diagrams in Fig.\,1  and after inclusion of the perturbation by 
the  test-particle it reads: 
\begin{eqnarray} {\rm Im}(B_0\!+\!B_1\!+\!B_2) &=& 4\pi^2 a \int\!\!{d^3 l \over 
(2\pi)^3}\,\delta(\vec l^{\,2}- \vec q^{\,2}) \bigg\{N_\eta(\vec P+\vec l, \vec p\,) 
\,N_\eta(\vec P-\vec l,\vec p\,)\nonumber \\ &&  +\Big[1-N_\eta(\vec P
+\vec l, \vec p\,)\Big]\Big[1-N_\eta(\vec P-\vec l, \vec p\,)\Big]\bigg\} 
\,.\end{eqnarray} 
Averaging again over the directions 
of $\vec p$, one finds to linear order in $\eta$:
\begin{equation}{\rm Im}(B_0\!+\!\bar B_1\!+\!\bar B_2)= a k_f \Big\{ I(s,\kappa) 
+ \tilde\eta \, \widehat I(s,\kappa,x)\Big\}\,, \end{equation} 
with the piecewise defined function:
\begin{equation} I(s,\kappa) = {1\over 2s}\, {\rm min}\Big(2s \kappa, 
|s^2+\kappa^2-1|\Big)\,. \end{equation} 
In order to arrive at this compact form one has consider separately the pertinent 
arrangement of three shifted spheres in the case $0<s<1$, where changes of 
$I(s,\kappa)$ occur at $\kappa = 1-s, \sqrt{1-s^2},\\1+s$, and in the case $s>1$, 
where changes of $I(s,\kappa)$ occur at $\kappa = s-1,s+1$. The correction term linear 
in $\tilde \eta$ is determined by the function:  
\begin{equation} \widehat I(s,\kappa,x) = {1\over s x}\, \theta(s+\kappa-x)
 \, \theta(x-|s-\kappa|)\,{\rm sign}(1+x^2-2(s^2+\kappa^2))\,, 
\end{equation}
\begin{equation} \widehat I(s,\kappa,0) = {2\over s}\,\delta(s-\kappa)\,
{\rm sign}(1-2s) \,. \end{equation}
The contribution of the diagram with two medium-insertion (i.e. the term in eq.(8) 
with two $N_\eta$-factors) is purely imaginary and after angular averaging it reads:
\begin{equation} \bar B_2 =2i a k_f \Big\{ I_*(s,\kappa) + \tilde\eta\, 
\widehat I_*(s,\kappa,x)\Big\}\,, \end{equation} 
with the function
\begin{equation} I_*(s,\kappa) =I(s,\kappa)\, \theta(1-s^2-\kappa^2) = 
\bigg\{\kappa \, \theta(1-s-\kappa) + {1\over 2s} (1-s^2-\kappa^2)\, 
\theta(s+\kappa-1) \bigg\}\,\theta(1-s^2-\kappa^2)\,,  \end{equation}
equal to the restriction of $I(s,\kappa)$ to the quarter unit disc $s^2+\kappa^2<1$, 
and the function
\begin{equation} \widehat I_*(s,\kappa,x) = {1\over s x}\, \theta(s+\kappa-x)
 \, \theta(x-|s-\kappa|)\,\theta(1+x^2-2(s^2+\kappa^2))\,, 
\end{equation}
\begin{equation} \widehat I_*(s,\kappa,0) = {2\over s}\,\delta(s-\kappa)\,
\theta(1-2s) \,, \end{equation}
equal to the restriction of $\widehat I(s,\kappa,x)$ to its domain of positive values.
A relation worth mentioning is: $\int_0^1 dx\, x^2\widehat I_*(s,\kappa,x) = 
2I_*(s,\kappa)$. For the sake of completeness we note that the vacuum term $B_0$ in 
dimensional regularization is simply: $B_0 = i a k_f  \kappa$. The support of the 
function $\widehat I_*(s,\kappa,x)$ in the $s\kappa$-plane is shown by the grey 
area in Fig.\,2. For $0<x<1$ this region is pieced together by a rectangle and a 
circular wedge. In the case $x>1$ only a circular wedge with radius 
$\sqrt{(1+x^2)/2}>1$ remains and the overlap of this circular wedge with the 
quarter unit disc, $s^2+\kappa^2<1$, vanishes for $x\geq \sqrt{2}$.   

\begin{figure}
\begin{center}
\includegraphics[scale=0.8,clip]{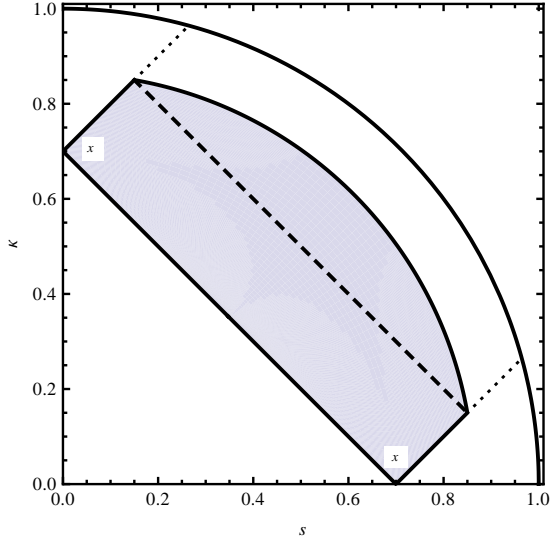}
\end{center}
\vspace{-.6cm}
\caption{For $0<x<1$ the support of the function $\widehat I_*(s,\kappa,x)$ 
in the $s\kappa$-plane consists of a rectangle and a circular wedge. The 
rectangle has side-lengths $\sqrt{2} x$ and $(1-x)/\sqrt{2}$, the 
radius of the boundary circle is $\sqrt{(1+x^2)/2}$. For $x>1$ this support 
consists only of  a circular wedge.}
\end{figure}

\subsection{Construction of the complex-valued single-particle potential}
Having available the real and imaginary part of the (on-shell) in-medium loop with 
corrections linear in $\tilde\eta$, the complex single-particle potential $U(p,k_f)+
i\,W(p,k_f)$ arising from ladder diagrams can be constructed in a one-step process 
from the expression for the resummed interaction density derived in ref.\cite{resum1}:
\begin{equation} V(\tilde\eta)= -{4\pi a\over M} \sum_{n=1}^\infty \sum_{j=0}^{n-1}
(B_0\!+\!\bar B_1)^{n-1-j} \bar B_2^{\,j}{n\!-\!1\choose j}{1\over j\!+\!1}= {4\pi 
a\over M \bar B_2} \, \ln{1-B_0-\bar B_1-\bar B_2 \over 1-B_0-\bar B_1}\,. 
\end{equation} 
The associated energy density is: ${\cal E}=(2\pi)^{-6} \int d^3p_1\int d^3p_2 \,
V(\tilde\eta)N_\eta(\vec p_1,\vec p\,)N_\eta(\vec p_2,\vec p\,)$.
There are two contributions to $U(p,k_f)+ i\,W(p,k_f)$ of different kinematical 
origin. The ``external'' contribution comes from the last two momentum-space 
integrations that are introduced by the closing of open ladder 
diagrams. The interesting term linear in $\eta$ leads to an integral over 
$\theta(k_f-|\vec p_1|)\,\delta^3(\vec p_2 -\vec p\,)$ times the resummed 
interaction density  $V(\tilde\eta)$ evaluated at $\tilde \eta=0$. The pertinent 
integral over a radius and a directional cosine can be transformed into the 
variables $(s,\kappa)$ and in the course of this transformation the function 
$s^2\kappa\,\widehat I_*(s,\kappa,x)$ arises as the appropriate weighting function. 
The ``internal'' contribution is an integral over $\theta(k_f-|\vec p_1|)\,
\theta(k_f-|\vec p_2|)$ (i.e. the product of two Fermi spheres) where the 
integrand is equal to the first-order expansion coefficient in $\tilde \eta$ of 
the resummed interaction density $V(\tilde\eta)$. The weighting function for this 
integral is $s^2 \kappa \,I_*(s,\kappa)$ (see eq.(15) in ref.\cite{resum1}). 
Since the external and internal contribution carry the same prefactor 
$2k_f^3/\pi^2$, the complex single-particle potential is completely determined by 
the quantity:  $\widehat I_*\, V(0)+ I_* \,V'(0)$. It has to be analyzed and 
decomposed into real and imaginary part on the different integration domains 
in the $s\kappa$-plane (i.e. the supports of $I_*(s,\kappa)$ and  $\widehat 
I_*(s,\kappa,x)$) separately for $0<x<1$ and $x>1$.  

Let us consider first the on-shell single-particle potential for momenta $p<k_f$ 
inside the Fermi sphere. After the cancellation of a term proportional to 
arctangent-function one arrives at the following concise double-integral 
representation for the real single-particle potential as derived from the 
resummed particle-particle and hole-hole ladder diagrams: 
\begin{eqnarray} U(p,k_f)&=& {8 a k_f^3\over M}\int\limits_0^1 \!ds\, s^2  
\!\!\int\limits_0^{\sqrt{1-s^2}}  \!\!d\kappa \, \kappa  \, 
\Bigg\{ {a k_f \widehat R(s,\kappa,x)I(s,\kappa)-\widehat I_*(s,\kappa, x) 
[\pi+ a k_f  R(s,\kappa)] \over [\pi+ a k_f R(s,\kappa)]^2+[ a k_f \pi\, 
I(s,\kappa)]^2}  \nonumber \\ & & \qquad \qquad \qquad \qquad  \qquad 
- {1\over a k_f}\widehat R(s,\kappa,x) \, \delta\bigg({\pi\over ak_f}+ 
R(s,\kappa)\bigg) \Bigg\}\,.
\end{eqnarray} 
The occurrence of the last $\delta$-function term in eq.(18) is very subtle. A
careful inspection of the mathematical expressions reveals that it is produced
by differentiating the discontinuity of the arctangent-function at infinity. 
The arctangent-function makes a jump by $-\pi$, when the denominator 
$\pi/ak_f+ R(s,\kappa) +\tilde \eta \, \widehat R(s,\kappa,x) $ of its argument 
passes through zero from positive to negative values. The $\delta$-function 
term\,\footnote{The 
$\delta$-function term affects also the calculation of the parameter $\zeta$ in 
section 6 of ref.\cite{resum1}. With the corrected integrand $I/(R^2+\pi^2 I^2)
- \delta(R)$ one obtains the (small) value $\zeta=0.1163$.} is treated numerically 
by first searching the line $\kappa(s)$ on which $R(s,\kappa(s))= -\pi/ak_f$ and 
then integrating over the appropriate  $s$-interval with the weighting function 
$|\partial R(s,\kappa)/\partial \kappa|^{-1}$. The line of zeros of the function 
$R(s,\kappa)$ inside the unit quarter disc is shown by the (lower) dashed-dotted 
line in Fig.\,3. This curve starts at $s_0=0$, $\kappa_0= 0.83356$ and ends at 
$s_1=0.55243$, $\kappa_1= 0.83356= \kappa_0$.  Alternatively, the $\delta$-function 
term can be treated in a regularized form such as: $\pi \,\delta(X) = 
\lim_{\epsilon\to 0^+} \epsilon/(X^2+\epsilon^2)$. One finds that finite $\epsilon 
\simeq 10^{-3}$ give sufficiently accurate numerical results.

\begin{figure}[ht]
\begin{center}
\includegraphics[scale=1.5,clip]{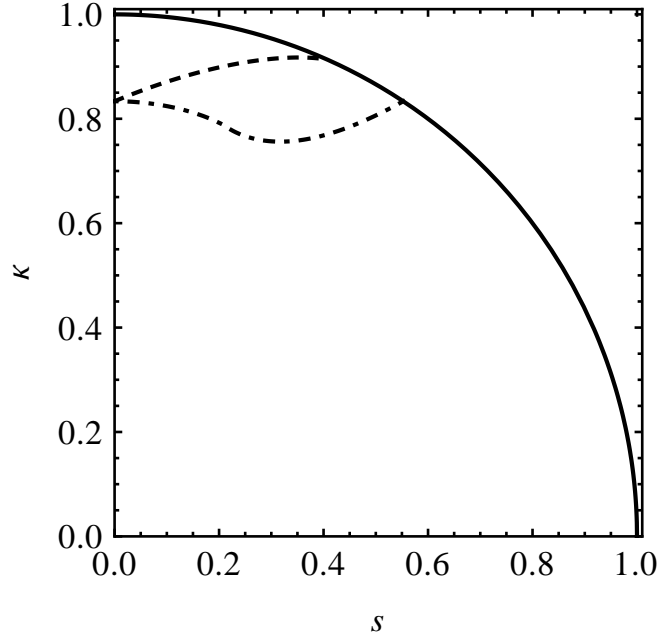}
\end{center}
\vspace{-.6cm}
\caption{The function $R(s,\kappa)$ defined in eq.(6) is zero along the (lower) 
dashed-dotted line. The function $F(s,\kappa)$  defined in eq.(40) is zero along 
the (upper) dashed line.}
\end{figure}

The imaginary single-particle potential for momenta $p<k_f$ inside the Fermi sphere 
is given by a simpler expression:
\begin{equation} W(p,k_f)= {8\pi a^2 k_f^4\over M}\int\limits_0^1 \!ds\, s^2  
\!\!\int\limits_0^{\sqrt{1-s^2}}  \!\!d\kappa \,\kappa \, {[\widehat I_*(s,\kappa,x)
-\widehat I(s,\kappa,x)]\, I(s,\kappa)\over [\pi+ a k_f R(s,\kappa)]^2
+[ a k_f \pi\, I(s,\kappa)]^2} \,.\end{equation}
Note that this integral receives contributions only from circular ring region 
indicated in Fig.\,2 by the dotted lines where the function $\widehat 
I(s,\kappa,x)$ is negative and $\widehat I_*(s,\kappa,x)$ vanishes. Consequently,  
$W(p,k_f)>0$ is positive for all $p<k_f$ as required for the stability of hole states 
in the Fermi sea. The important constraint $W(k_f,k_f)=0$ (Luttinger's theorem 
\cite{luttinger}) is also obvious from the representation in eq.(19), since at 
$x=1$ the integration region in the $s\kappa$-plane vanishes identically.

\subsection{Continuation into the region outside the Fermi sphere}
Next, we consider the complex single-particle potential for momenta $p>k_f$ outside 
the Fermi sphere. The new element here is that the pertinent integration region 
in the $s\kappa$-plane is now larger than the quarter unit disc.
The contribution to the real single-particle potential $U(p,k_f)$ from the
region  $s^2+\kappa^2<1$ is of the same form as described in subsection 2.2. 
The corresponding integral will include the weighting function $\theta(1-s^2-
\kappa^2)$. The ``external'' contribution from the region $s^2+\kappa^2>1$ requires 
to take the limit $\tilde \eta\to 0$ because of the factor $1/\tilde\eta$ which 
comes from $1/\bar B_2$ in $V(\tilde \eta)$, when $I_*(s,\kappa)=0$. In effect the  
``external'' piece continues the term proportional to $\widehat I_*(s,\kappa,x)$ 
that is already present in the ``internal'' piece into the region $s^2+\kappa^2>1$. 
The complete expression for the real single-particle potential $U(p,k_f)$ outside 
the Fermi sphere $p>k_f$ reads: 
\begin{eqnarray} U(p,k_f)&\!\!\!\!=\!\!\!\!& {8 a k_f^3\over M}\int
\limits_0^{(x+1)/2} \!ds\, s^2  \!\!\int\limits_0^{(x+1)/2} \!\!d\kappa \,\kappa\, 
\Bigg\{ {a k_f \widehat R(s,\kappa,x)I_*(s,\kappa)-\widehat I_*(s,\kappa, x) 
[\pi+ a k_f  R(s,\kappa)] \over [\pi+ a k_f R(s,\kappa)]^2+[ a k_f \pi\, 
I(s,\kappa)]^2}  \nonumber \\ & & \qquad \qquad \qquad \qquad  \qquad  \qquad 
-\, {\theta(1-s^2-\kappa^2)\over a k_f}\widehat R(s,\kappa,x) \, \delta
\bigg({\pi\over ak_f}+ R(s,\kappa)\bigg) \Bigg\}\,.\end{eqnarray}
Note that the square of side-length $(x+1)/2>1$ is large enough to cover the 
relevant integration regions (the quarter unit disc and the circular wedge of 
radius $\sqrt{(1+x^2)/2}>1$) in the $s\kappa$-plane. 

The calculation of the ``external'' contribution from the region $s^2+\kappa^2>1$ 
through taking the limit $\tilde \eta\to 0$ gives rise to a complex-valued 
single-particle potential, whose imaginary part reads: 
\begin{equation} W(p,k_f)= -{8\pi a^2 k_f^4\over M}\int\limits_0^{(x+1)/2} \!ds\, 
s^2 \!\!\int\limits_0^{(x+1)/2}  \!\!d\kappa \,\kappa \, {\widehat I_*(s,\kappa,x) 
I(s,\kappa)\,\theta(s^2+\kappa^2-1)\over [\pi+ a k_f R(s,\kappa)]^2
+[ a k_f \pi\, I(s,\kappa)]^2} \,.\end{equation}
Note that for $x>\sqrt{2}$ the condition $s^2+\kappa^2>1$ can be dropped since then 
the support of the weighting function $ \widehat I_*(s,\kappa,x)$ (a circular wedge)
lies completely in the outer region. The integrand in eq.(21) is obviously 
positive and therefore $W(p,k_f)<0$ is negative for all $p>k_f$ as required for the 
stability of particle states outside the Fermi sphere. 

\subsection{Hugenholtz-Van-Hove theorem}
An important constraint on the single-particle potential is given by the 
Hugenholtz-Van-Hove theorem \cite{hugenholtz}. It states that the total 
single-particle energy $U(k_f,k_f)+k_f^2/2M$ at the Fermi surface $p= k_f$ is 
equal to the chemical potential. By a general thermodynamical relation the 
chemical potential is equal to the derivative of the energy density 
$\rho\, (\bar E(k_f)+3k_f^2/10M)$ with respect to the particle density 
$\rho = k_f^3/3\pi^2$. We prove now that the Hugenholtz-van-Hove theorem is 
satisfied in the present non-perturbative calculation. The starting point is the 
resummed (interaction) energy per particle $\bar E(k_f)$ as given by eq.(14) in 
ref.\cite{resum1}. First, one calculates the derivative with the respect to $k_f$ in 
that given double-integral representation. Then, one remembers that the interaction
energy density was originally an integral over the product of two Fermi 
spheres, $\theta(k_f-|\vec p_1|)\,\theta(k_f-|\vec p_2|)$, and differentiates 
separately the integration boundaries and the integrand with respect to $k_f$. 
The described procedure translates into the following sequence of equations:     
\begin{eqnarray} && \bar E(k_f) +{k_f \over 3} {\partial \bar 
E(k_f) \over \partial k_f} =-{8 k_f^2 \over \pi M } 
\int\limits_0^1\!ds\,s^2  \!\!\int\limits_0^{\sqrt{1-s^2}} \!\!d\kappa \,\Bigg\{ 
5 \arctan{ ak_f I(s,\kappa)\over 1+ ak_f \pi^{-1}R(s,\kappa)} \nonumber \\ && 
+{a k_f I(s,\kappa) \over [1+ a k_f \pi^{-1} R(s,\kappa)]^2+[ a k_f I(s,\kappa)
]^2} -{\pi^2\over ak_f}\,  \delta\bigg({\pi \over ak_f}+ R(s,\kappa)\bigg) 
\Bigg\} \nonumber \\ &&  = {8 a k_f^3\over M}\int\limits_0^1 \!ds\, 
s^2  \!\!\int\limits_0^{\sqrt{1-s^2}}  \!\!d\kappa \, \kappa  \,\Bigg\{{a k_f 
\widehat R(s,\kappa,1)I(s,\kappa)-\widehat I_*(s,\kappa, 1) [\pi+ a k_f  
R(s,\kappa)] \over [\pi+ a k_f R(s,\kappa)]^2+[ a k_f \pi\, I(s,\kappa)]^2} 
\nonumber \\ && \hspace{4.5cm} - {1\over ak_f}\widehat R(s,\kappa,1) \,\delta
\bigg({\pi \over ak_f}+ R(s,\kappa)\bigg) \Bigg\} = U(k_f,k_f) \,,\end{eqnarray}
where the equality between the initial and final term is precisely the statement 
of the Hugenholtz-Van-Hove theorem \cite{hugenholtz}. In order to establish the 
agreement with  $U(k_f,k_f)$ the following identities have been instrumental:
\begin{equation} {\partial \over \partial k_f} \Big[ k_f R(P/k_f,q/k_f) \Big] 
= {1\over s}\ln{(s+1)^2-\kappa^2\over |(s-1)^2-\kappa^2|} = \widehat R(s,\kappa,1) 
\,, \end{equation} 
\begin{equation} {\partial \over \partial k_f} \Big[ k_f I(P/k_f,q/k_f) \Big]= 
{1\over s}\, \theta(s+\kappa-1) = \widehat I_*(s,\kappa,1) \,, \end{equation} 
where the derivative with respect to $k_f$ is taken at fixed $P$ and $q$, and the 
constraint $s^2+\kappa^2<1$ holds here. Note that the arctangent-function 
in eq.(22) refers to the usual branch with odd parity, $\arctan(-X) = - 
\arctan X$, and values in the interval $[-\pi/2, \pi/2]$.  Other branches of the 
arctangent-function are excluded by the weak coupling limit $a\to 0$, which has 
to give zero independently of the sign of the scattering length $a$. The 
$\delta$-function terms in both double-integral expressions in eq.(22) are crucial 
for the actual numerical validity of the Hugenholtz-Van-Hove theorem.  We have 
examined this over a wide range of positive and negative values of the 
dimensionless coupling strength $ak_f$. In fact the violation of the 
Hugenholtz-Van-Hove theorem without the $\delta$-function terms has given the 
hint that there is this subtlety in differentiating the arctangent-function (as 
explained in the subsection 2.2). 

The slope of the (on-shell) single-particle potential at the Fermi surface $p=k_f$ 
determines the density-dependent effective mass $M^*(k_f)$ of the (stable) 
quasi-particle excitations at the Fermi surface. The corresponding relation for 
the effective mass $M^*(k_f)$ reads:
\begin{equation}  {1\over M^*(k_f)} = {1\over M} +{1\over k_f} \, 
{\partial U(p,k_f) \over \partial p}\bigg|_{p=k_f} \,, \end{equation} 
with $M$ the free fermion mass.
\subsection{Perturbative expansion}
Given the closed-form expressions for the complex single-particle potential 
$U(p,k_f)+i\,W(p,k_f) $ in eqs.(18-21), one can expand them in powers of the 
scattering length $a$. It is convenient to scale out the free Fermi energy 
$k_f^2/2M$ and to use a dimensionless expansion parameter, such that the 
perturbative series reads:  
\begin{equation} U(p,k_f)+i\,W(p,k_f)= {k_f^2 \over 2M} \sum_{n=1}^\infty (-a k_f)^n  
\Big[\Phi_n(x)+i\,\Omega_n(x)\Big]\,.  \end{equation} 
Here, the dimensionless functions  $\Phi_n(x)$ and $\Omega_n(x)$ describe the 
momentum dependence of the real and imaginary part at $n$-th order. Note that the 
$\delta$-function term in eqs.(18,20) proportional to $\delta(\pi+ ak_f 
R(s,\kappa))$ does not contribute at any order in the expansion in powers of 
$-ak_f$. In this sense it represents a truly non-perturbative term. 
The  first order contribution to the perturbative expansion in eq.(26) is the 
trivial Hartree-Fock mean field result:
\begin{equation} \Phi_1(x)={ 4\over 3\pi}\,,  \qquad \Omega_1(x)=0\,,\end{equation}
and the second order contribution to the complex single-particle potential is
known from the classical work by Galitskii \cite{galitski}. The corresponding 
function $\Phi_2(x)$ has for $0<x<1$ the form:  
\begin{equation} \Phi_2(x)={ 4\over 15\pi^2}\Bigg\{ 11-2x^4 \ln{1-x^2\over x^2} 
+{10\over x}(1-x^2) \ln{1+x\over 1-x} -{2\over x} 
(2-x^2)^{5/2}  \ln{1+x \sqrt{2-x^2} \over 1-x^2}\Bigg\} \,,\end{equation}
and it gets continued into the region $x>1$ by the expression: 
\begin{eqnarray} \Phi_2(x)&=&{ 4\over 15\pi^2}\Bigg\{ 11-2x^4 \ln{x^2-1\over 
x^2} +{10\over x}(1-x^2) \ln{x+1\over x-1} -{2\over x} \bigg[ \theta(
\sqrt{2}-x)\,\nonumber \\ && \times  (2-x^2)^{5/2}  \ln{1+x \sqrt{2-x^2} \over 
x^2-1}+ \theta(x-\sqrt{2})\, (x^2-2)^{5/2}  \arcsin{1\over x^2-1}\bigg] \Bigg\} 
\,.\end{eqnarray}
The function $ \Omega_2(x)$ describing the leading contribution to the 
imaginary potential has the form:
\begin{equation} \Omega_2(x)= {\theta(1-x) \over 2\pi}\,(1-x^2)^2+ {2\,\theta(x-1)
\over 15\pi x}\, \Big\{7-5x^2-2(2-x^2)^{5/2}\,\theta(\sqrt{2}-x)\Big\} \,.
\end{equation}
The elaborate function $\Phi_2(x)$ has the boundary values $\Phi_2(0)=4/\pi^2$,
$\Phi_2(1)=4(11-2\ln 2)/15\pi^2$ and the slope $\Phi_2'(1)=16(1-7\ln 2)/15\pi^2$ 
at $x=1$. A good approximation of $\Phi_2(x)$ for $x>5/3$ is provided by 
the asymptotic expansion: $\pi^2 \Phi_2(x)= (16/9x^2)+ (22/45x^4)+ (32/105x^6)+ 
{\cal O}(x^{-8})$.

\begin{figure}[ht]
\begin{center}
\includegraphics[scale=0.4,clip]{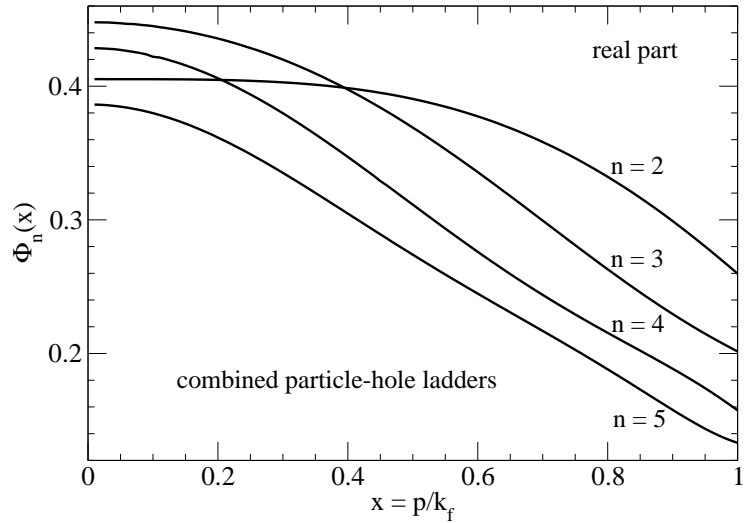}
\end{center}
\vspace{-.8cm}
\caption{Momentum dependence of perturbative contributions to the real 
single-particle potential inside the Fermi sphere.}
\end{figure}

The agreement at first order follows from the value of the integral: $\int\!ds 
\,s^2\!\int\!d\kappa\, \kappa\,\widehat I_*(s,\kappa, x)=1/12$, which applies both 
for $0<x<1$ and $x>1$. In the same way, the double-integral representations of 
$\Phi_2(x)$ and $\Omega_2(x)$ which one obtains by expanding eqs.(18-21) to second 
order in $ak_f$ are in perfect numerical agreement (at the 6-digit level) with the 
analytical expressions written in eqs.(28-30), both for $0<x<1$ and $x>1$. This 
agreement provides a very important check on the formalism introduced in subsection 2.1 
in order to construct non-perturbatively the complex single-particle potential 
$U(p,k_f)+i\, W(p,k_f)$. While results for the interaction energy per particle 
$\bar E(k_f)$ are available up to fourth order \cite{hammer,steele}, the on-shell 
single-particle potential $U(p,k_f)+i\, W(p,k_f)$ has so far not been computed beyond 
second order. At this point the present calculation provides a multitude of new 
results. One should note that at third and higher order in $ak_f$ there exist also 
other classes of diagrams, such as the particle-hole ring diagrams \cite{schaefer}. 
The corresponding real single-particle potential $U(p,k_f)^{n-\rm ring}$ is studied 
in section 4.

\begin{figure}[ht]
\begin{center}
\includegraphics[scale=0.4,clip]{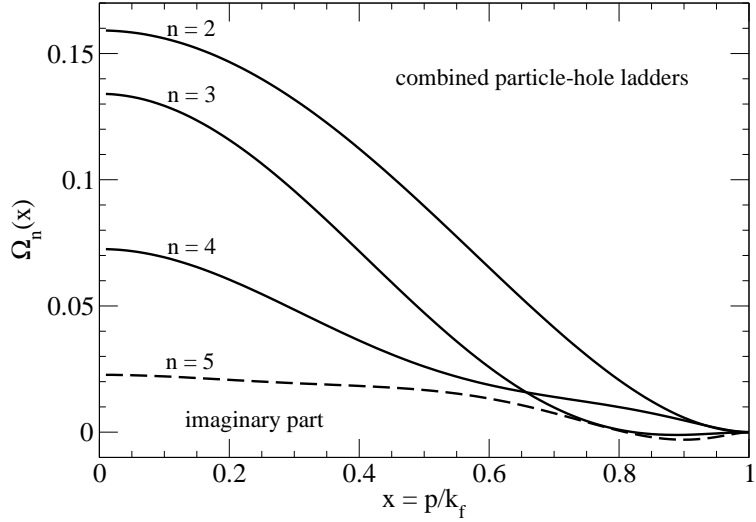}
\end{center}
\vspace{-.8cm}
\caption{Momentum dependence of perturbative contributions to the imaginary 
single-particle potential inside the Fermi sphere.}
\end{figure}

The $x$-dependence of the functions $\Phi_n(x)$ and $\Omega_n(x)$ for $n=2,3,4,5$ 
in the region $0<x<1$ is shown in Figs.\,4,5. One observes a decrease of the 
$n$-th order potential-functions as one moves from $x=0$ (bottom of 
the Fermi sea) to $x=1$ (at the Fermi surface). There is also a tendency that 
higher-order functions get smaller in magnitude. Precise numerical values of the 
boundary values $\Phi_n(0)$, $\Omega_n(0)$  and $\Phi_n(1)$ are listed in Table\,1 
up to order $n=8$. The values $\Phi_n(0)$ and $\Omega_n(0)$ have been computed by 
using the explicit expressions for $\widehat R(s,\kappa,0)$, $\widehat 
I(s,\kappa,0)$ and  $\widehat I_*(s,\kappa,0)$ written in eqs.(7,12,16). The 
analytical value $\Omega_3(0)= 16[2 \sqrt{2}\ln(1+\sqrt{2})-3 \ln 2]/5\pi^2$ 
associated to the imaginary single-particle potential at third order is obtained 
as a byproduct of this calculation. The right boundary values $\Phi_n(1)$ are 
furthermore determined by the Hugenholtz-Van-Hove theorem. One has the relation 
$\Phi_n(1) = (2/\pi)^n (n+5)\, c_n/9$,  where the coefficients $c_n$ have been 
defined through the $k_f$-expansion of $\bar E(k_f)$ in eq.(19) of ref.\cite{resum1}. 
Inserting the numerical values of $c_n$ listed in eq.(20) of ref.\cite{resum1} one 
finds perfect agreement with the values of $\Phi_n(1)$ calculated directly  
via the perturbative expansion of $U(p,k_f)$. This demonstrates that the 
Hugenholtz-Van-Hove theorem is fulfilled order by order without the 
$\delta$-function term in eq.(18). 

\begin{table}
\begin{center}
\begin{tabular}{|c|c c c c c c c c|}\hline
$n$ & 1 & 2 & 3 & 4 & 5& 6 & 7& 8 \\ \hline 
$\Phi_n(0)$ &  0.42441  & 0.40528 & 0.44788 & 
0.42857 & 0.38638 & 0.33254 & 0.28779 & 0.25073 \\ \hline 
$\Omega_n(0)$ &    & 0.15915 & 0.13406  & 
0.07256 & 0.02273 & --0.00496 & --0.01414 & --0.01237 \\ \hline 
$\Phi_n(1)$ &  0.42441  & 0.25975 & 0.20153 & 
0.15744 & 0.13314 & 0.11153 &0.09959 & 0.08573\\ \hline\hline 
$\Psi_n(0)$ &  0.42441  &  0.40528 & 0.35745 & 0.29486 & 0.23203  &  0.17676
& 0.13168  & 0.09658 \\ \hline
$\Xi_n(0)$ &    & 0.15915 & 0.16809 & 0.13425  & 
0.09601 & 0.06479 & 0.04222 & 0.02689 \\ \hline 
$\Psi_n(1)$ &  0.42441  & 0.25975 & 0.17083 &0.11493 &0.07872 & 0.05444 
&  0.03797 & 0.02663\\ \hline
\end{tabular}
\caption{Boundary values of the functions $\Phi_n(x)$,  $\Omega_n(x)$,
$\Psi_n(x)$ and $\Xi_n(x)$ describing the momentum dependence of perturbative 
contributions to the complex single-particle potential. The values $\Phi_n(1)$ and 
$\Psi_n(1)$ are fixed by the Hugenholtz-Van-Hove theorem.}
\end{center}
\end{table}

The continuation of the potential-functions $\Phi_n(x)$ and $\Omega_n(x)$ for 
$n=2,3,4,5$ into the region $1<x<2$ outside the Fermi sphere is shown in 
Figs.\,6,7. The behavior with increasing $x$ is nonuniform and alternates for 
different orders including even changes of the sign. Note that a small positive 
contribution to the imaginary single-particle potential $W(p,k_f)$ for $p>k_f$ at 
higher orders in the perturbative expansion is compatible with the stability 
(i.e. damping) of quasi-particle excitations.

\begin{figure}[h]
\begin{center}
\includegraphics[scale=0.4,clip]{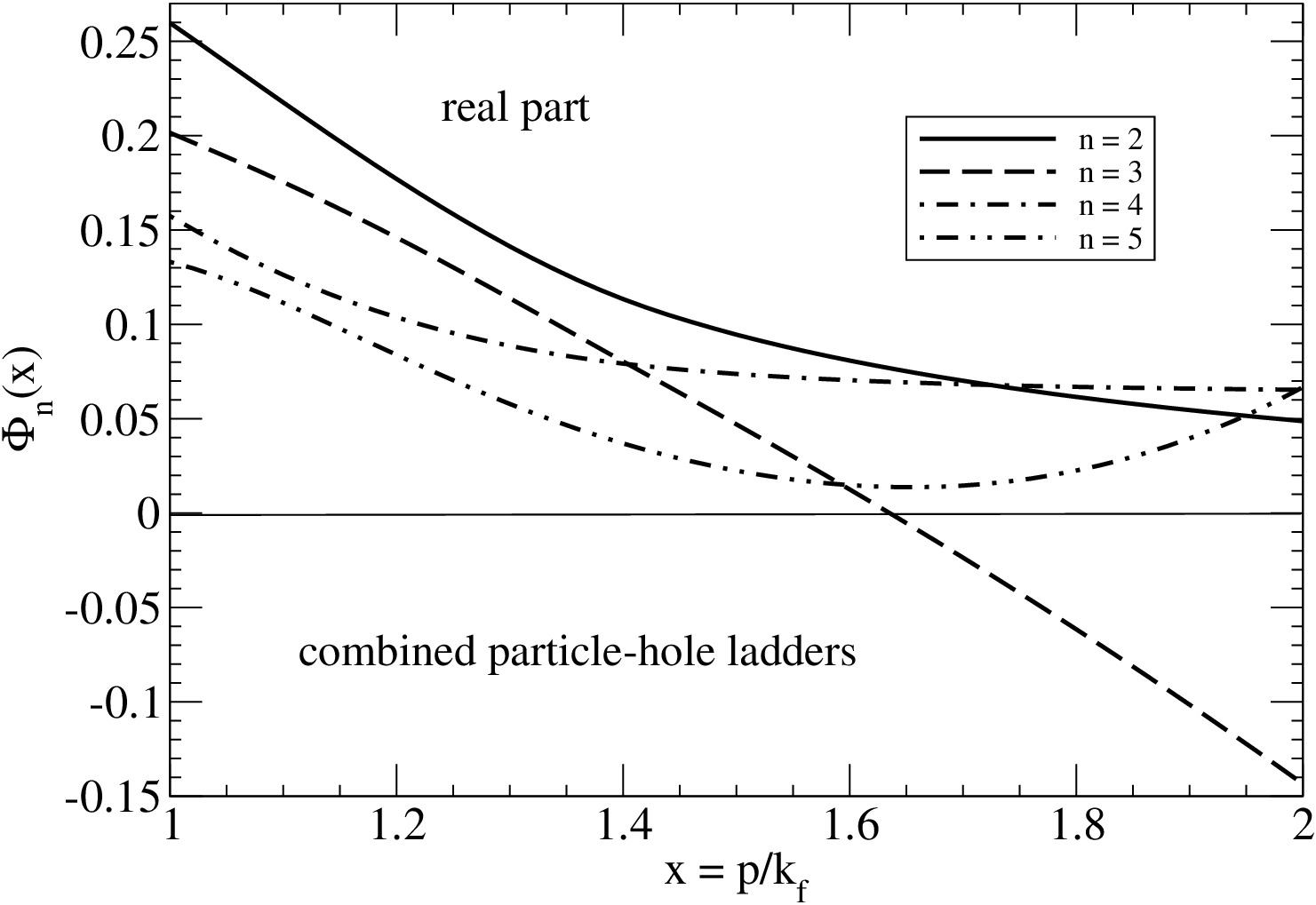}
\end{center}
\vspace{-.8cm}
\caption{Momentum dependence of perturbative contributions to the real 
single-particle potential outside the Fermi sphere.}
\end{figure}

\begin{figure}[h]
\begin{center}
\includegraphics[scale=0.4,clip]{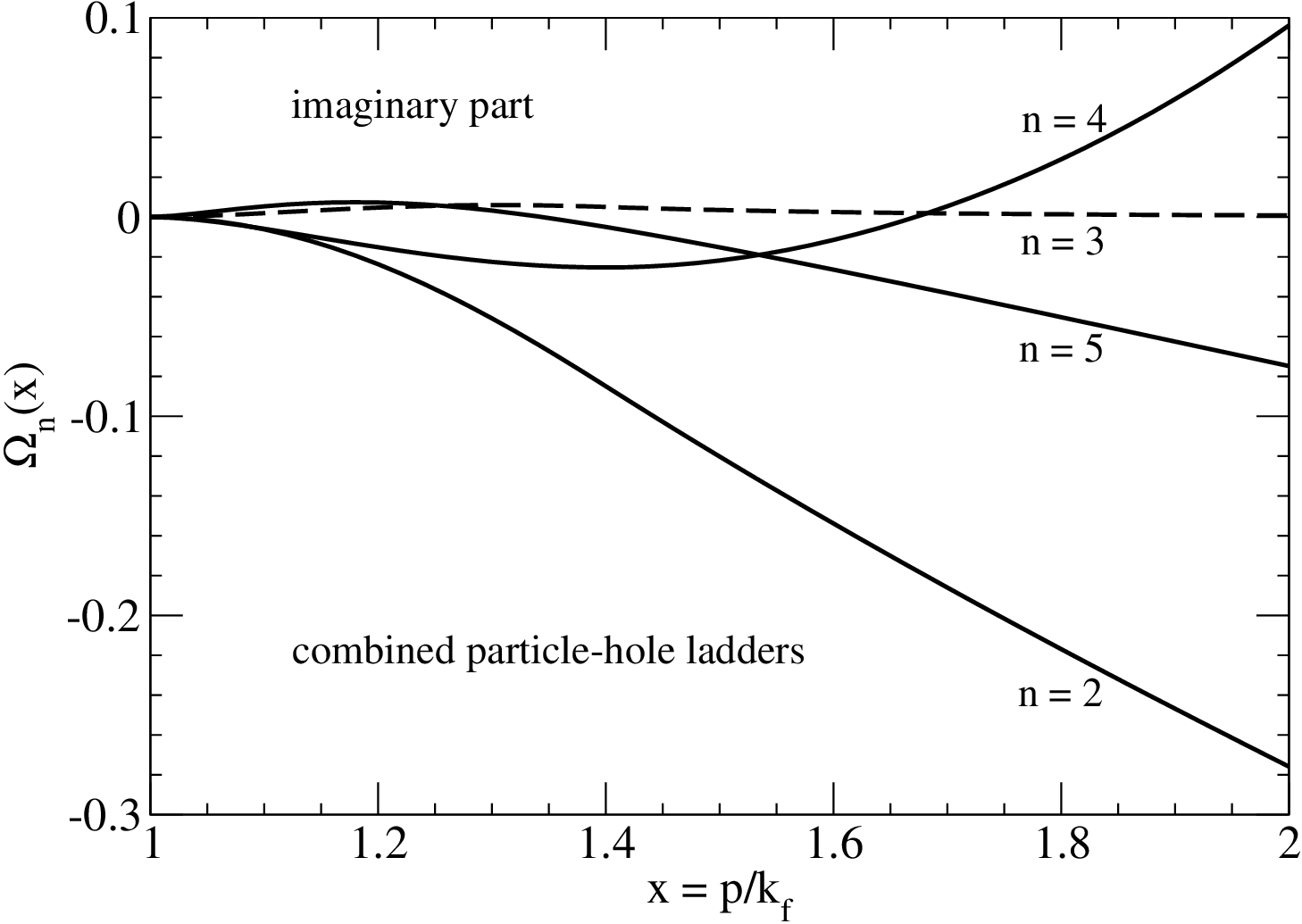}
\end{center}
\vspace{-.8cm}
\caption{Momentum dependence of perturbative contributions to the imaginary 
single-particle potential outside the Fermi sphere.}
\end{figure}

\subsection{Strong coupling limit} 
The limit $a\to \infty $ is of special interest since in this limit the 
strongly interacting many-fermion system becomes scale invariant. Returning to the 
expressions for the resummed single-particle potential $U(p,k_f)+i\, W(p,k_f)$ in 
eqs.(18-21), one sees that the limit $a\to -\infty $ (in the repulsive 
regime) can be performed straightforwardly.  The resulting real and imaginary part 
read for $p<k_f$: 
\begin{eqnarray} U(p,k_f)^{(\infty)} &\!\!\!=\!\!\!& {8 k_f^2\over M}
\int\limits_0^1 \!ds\, s^2  \!\!\int\limits_0^{\sqrt{1-s^2}}  \!\!d\kappa \, 
\kappa  \,\Bigg\{{\widehat R(s,\kappa,x)I(s,\kappa)-\widehat I_*(s,\kappa, x) 
R(s,\kappa) \over R(s,\kappa)^2+\pi^2 I(s,\kappa)^2} -\widehat R(s,\kappa,x) 
\,\delta(R(s,\kappa)) \Bigg\} \nonumber \\ &\!\!\!=\!\!\!& 
{k_f^2 \over 2M}\, \Phi_{\rm uni}(x) \,, \end{eqnarray}
\begin{equation} W(p,k_f)^{(\infty)} = {8 \pi k_f^2\over M}
\int\limits_0^1 \!ds\, s^2  \!\!\int\limits_0^{\sqrt{1-s^2}}  \!\!d\kappa \, 
\kappa  \,{[\widehat I_*(s,\kappa,x)-\widehat I(s,\kappa, x)] \,I(s,\kappa) 
\over R(s,\kappa)^2+\pi^2 I(s,\kappa)^2}= {k_f^2 \over 2M}\, \Omega_{\rm uni}(x) 
\,, \end{equation}
and the modifications of these formulas for $p>k_f$ are obvious from eqs.(20,21). The 
dimensionless functions $\Phi_{\rm uni}(x)$ and $\Omega_{\rm uni}(x)$ describe the 
dependence on the rescaled momentum variable $x=p/k_f$. The calculated boundary 
values are:
\begin{equation} \Phi_{\rm uni}(0)= -0.214\,, \qquad \Phi_{\rm uni}(1)= \xi_n-1 
=-0.493\,,  \qquad \Omega_{\rm uni}(0)=0.780\,, \end{equation} 
with $\xi_n=0.507$ the normal Bertsch parameter obtained in ref.\cite{resum1}. Note 
that the value $\Phi_{\rm uni}(1)= \xi_n-1$ is fixed by the Hugenholtz-Van-Hove 
theorem  and interestingly it is almost entirely determined by the contribution of 
the $\delta$-function term in eq.(31): $\Phi_{\rm uni}(1)=-0.031-0.462$. Given the 
two negative boundary values in eq.(33) one would expect a monotonically decreasing 
function $\Phi_{\rm uni}(x)$. However, the numerical evaluation of the full expression 
for $\Phi_{\rm uni}(x)$ in eq.(31) leads to a completely different result, which is 
shown by the full line in Fig.\,8. The function $\Phi_{\rm uni}(x)$ rises 
steeply into the positive domain, it develops a peak of height $0.98$ at 
$x \simeq 0.3$, and then drops back to negative values. The negative slope of 
$\Phi_{\rm uni}(x)$ at $x=1$ translates into an enhanced effective mass
\footnote{This is very different 
from the quantum Monte-Carlo result $M^*/M \simeq 1$ of ref.\cite{mag}.} 
of $M^*/M=[1+\Phi'_{\rm uni}(1)/2]^{-1} \simeq 2.1$. The average value
of $ \Phi_{\rm uni}(x)$ taken over the interior of the Fermi sphere is: $3\int_0^1 
dx\,x^2 \Phi_{\rm uni}(x) = -48 \int ds\,s^2 \int d\kappa \,\kappa\, I_* 
R/(R^2+\pi^2 I^2) = -0.224$. This is rather close to the value $\Phi_{\rm uni}(0) = 
-0.214$ at the bottom of the Fermi sea. Furthermore, one observes that the function 
$\Omega_{\rm uni}(x)$ associated to the imaginary single-particle potential 
$W(p,k_f)^{(\infty)}$ follows the up- and downward motion of $\Phi_{\rm uni}(x)$. This 
feature implies that the peculiar unbound hole excitations 
with momenta $0.2k_f< p< 0.5 k_f$ are at the same time also very short-lived. 

The strong momentum dependence of the potential $U(p,k_f)^{(\infty)}$ indicates 
an instability against a (topological) phase transition to a state with separation in 
momentum space \cite{sarma,volovik,pankrat}. Namely, if one adds the kinetic energy 
$p^2/2M$ to the potential $U(p,k_f)^{(\infty)}$, one finds a momentum region where the full 
single-particle energy lies above the (interacting) Fermi 
energy $\xi_n k_f^2/2M$. In Fig.\,13 this corresponds to the interval $0.272<x<0.415$, 
inside which $\Phi_{\rm uni}(x)>\xi_n-x^2$ holds. But states lying above the Fermi energy 
will not be occupied, so one gets an empty shell (or bubble) in the Fermi sea, which 
reduces the density by about $5\%$ to $0.95\, k_f^3/3\pi^2$. Such a bubble formation in 
neutron matter (which could have important consequences for neutron star cooling) has 
been discussed in ref.\cite{pankrat} in connection with the possibility of 
neutral pion condensation. According to the present calculation the low-energy 
neutron-neutron interaction,  as represented 
by the large $nn$-scattering length $a_{nn}= 19\,$fm, could likewise induce the 
instability to this (topological) phase transition. When analyzing the potential 
$U(p,k_f)$ in eq.(18) at finite coupling strength $ak_f$, one finds in a numerical 
study that the instability occurs for $ak_f>5.5$ and  $ak_f<-3.0$.   

\begin{figure}[t]
\begin{center}
\includegraphics[scale=0.4,clip]{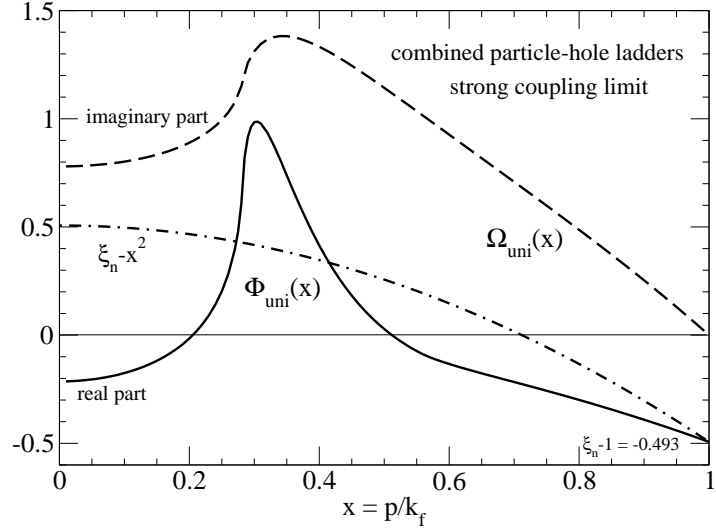}
\end{center}
\vspace{-.8cm}
\caption{Momentum dependence of the single-particle potential inside the Fermi 
sphere in the limit $a\to \infty$.}
\end{figure}
\begin{figure}[h]
\begin{center}
\includegraphics[scale=0.4,clip]{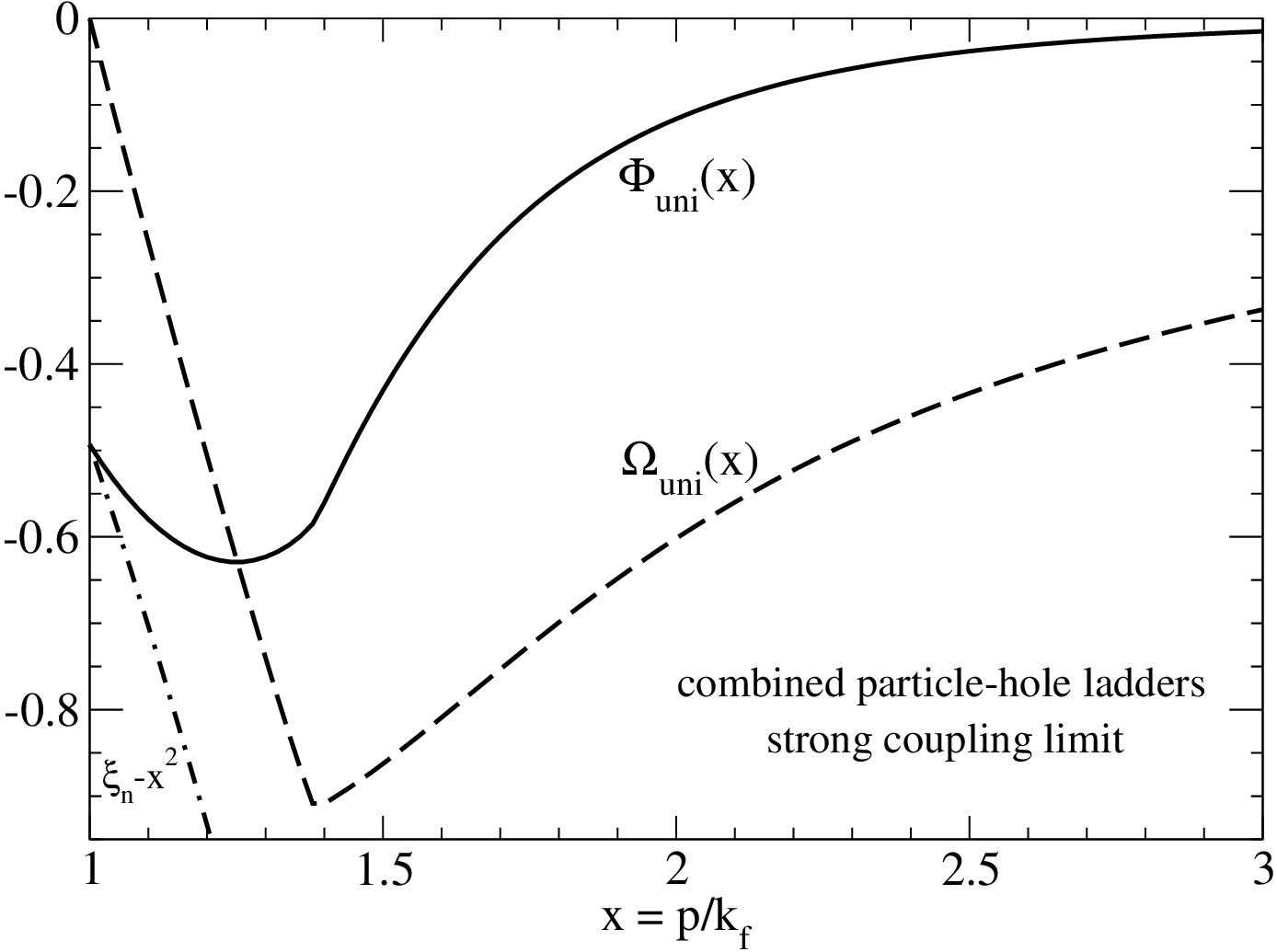}
\end{center}
\vspace{-.8cm}
\caption{Momentum dependence of the single-particle 
potential outside the Fermi sphere in the limit $a\to \infty$. }
\end{figure}

The continuation of the functions $\Phi_{\rm uni}(x)$ and $\Omega_{\rm uni}(x)$ into the 
region $x>1$ outside the Fermi surface is shown in Fig.\,9. One observes that 
$\Phi_{\rm uni}(x)$ reaches its minimum value of $-0.629$ at $x \simeq 1.25$ and from 
there on it decreases rapidly in magnitude with increasing $x= p/k_f$. The other 
function $\Omega_{\rm uni}(x)$ drops linearly from zero at $x=1$ to its minimum value 
of $-0.909$ at $x\simeq 1.4$ and from there on it decreases slowly in magnitude with 
increasing $x$. This means that particle excitations with relatively high 
momentum get weakly attracted by fermionic medium and are at the same time very 
short-lived. The asymptotic behavior of the function $\Omega_{\rm uni}(x)$ for large $x$ is:
\begin{equation} \Omega_{\rm uni}(x) = -{8 \over 3\pi x}\bigg( 1+{8\over 5x^2} 
+{4\over 3x^4} -{1\over x^6} +\dots \bigg)\,, \end{equation}
with coefficients determined in an extensive numerical study. The reason for such
a simple asymptotic form is the integration region in eq.(21) which becomes for 
large $x$ a very thin circular wedge spanned between the points $((x-1)/2,(x+1)/2)$ 
and $((x+1)/2,(x-1)/2)$ in the $s\kappa$-plane. In this situation simplifying 
approximations will hold for the functions $R(s,\kappa)$ and $I(s,\kappa)$ in 
the integrand.

Another special feature visible from Figs.\,8,9 is that the imaginary 
part $\Omega_{\rm uni}(x)$ vanishes linearly at 
the Fermi surface $x=1$, with the same slope  $\Omega_{\rm uni}'(1) =-4\pi(s_1\kappa_1)^2= 
-2.6647$ from both sides. Indeed, when evaluating numerically $W(p,k_f)$ in eqs.(19,21) 
at small coupling strength $|a k_f|< 1/2$ one finds the quadratic behavior 
$\pm (1-x)^2$ according to Luttinger's theorem \cite{luttinger}. However, for 
large coupling this degenerates to a linear behavior. Let us point to the 
origin of this unconventional property. The numerator in the representations of 
$W(p,k_f)$ in eqs.(19,21) alone would lead to the quadratic behavior $\pm(1-x)^2$,
but there is also the denominator (due to resummation) which introduces a
singular region into the double-integral. In principle the linear behavior $1-x$, 
which is caused by the vanishing of $R(s,\kappa)+\pi/ak_f$ on the unit-circle 
$s^2+\kappa^2=1$, is present for all coupling strengths $ak_f>0$ and $ak_f<-\pi/2$.
Using the dimensionless function $\Omega(x)= 2M k_f^{-2}\,W(p,k_f)$, one finds in 
a careful analysis the following result for the (negative) slope at the Fermi 
surface $x=1$:
\begin{equation} \Omega'(1) = -{8\pi s_1^2(1-s_1^2) \over 2+\pi(ak_f)^{-1}s_1^2}
\,, \end{equation}
where $0<s_1<1$ is the solution of the equation $R(s_1,\sqrt{1-s_1^2})= -\pi/ak_f$.
Interestingly, the largest negative slope $\Omega'(1) \simeq -2\pi$ is not reached 
at infinite coupling, but for $ak_f \simeq -\pi/2$ (approached from below).

\section{Particle-particle ladder diagrams only}
In this section we perform the analogous construction of the complex single-particle 
potential $U(p,k_f)+i\,W(p,k_f)$ for the particle-particle ladder diagrams. This 
subclass of diagrams can be summed to all orders in the form of a geometrical 
series \cite{schaefer}. The momentum regions $p<k_f$ (inside the Fermi sphere) and 
$p>k_f$ (outside the Fermi surface) need to be considered separately.

The starting point is again the pertinent in-medium loop (also called 
particle-particle bubble function \cite{schaefer}). With inclusion of the 
test-particle with momentum $\vec p$ (see eq.(2)), the on-shell in-medium loop 
takes now the form: 
\begin{equation} B_p = 4\pi a \int\!\!{d^3 l \over (2\pi)^3}\,{1\over \vec 
l^{\,2}- \vec q^{\,2}-i\epsilon} \Big[1-N_\eta(\vec P+\vec l,\vec p\,)\Big] 
\Big[1-N_\eta(\vec P-\vec l,\vec p\,)\Big]\,,\end{equation}
where each factor $1-N_\eta(\dots)$ describes a particle phase-space.  
Averaging again over the directions of $\vec p$, one gets to linear order in the 
parameter
$\tilde\eta= \eta\,\pi^2/k_f^3$:
\begin{equation} {\rm Re}\,\bar B_p= -{a k_f \over \pi} \Big\{ F(s,\kappa) + 
\tilde\eta \, \widehat F(s,\kappa,x)\Big\}\,, \end{equation} 
\begin{equation} {\rm Im}\,\bar B_p= a k_f  \Big\{ I(s,\kappa) -I_*(s,\kappa) + 
\tilde\eta\,  \Big[\widehat I(s,\kappa,x)-\widehat I_*(s,\kappa,x)\Big]\Big\}\,, 
\end{equation} 
with the functions  $I(s,\kappa) -I_*(s,\kappa)$ and $\widehat I(s,\kappa,x)-
\widehat I_*(s,\kappa,x)$ determined by the analytical expressions written in 
eqs.(10-16). The interaction density of the resummed particle-particle ladder 
diagrams is a geometrical series and it has the form:
\begin{equation} V_p(\tilde \eta) = -{4\pi a \over M}\sum_{n=1}^\infty \bar B_p^{\,n-1} = 
{4\pi a \over M(\bar B_p-1)}\,.\end{equation} 
\subsection{Single-particle potential inside the Fermi sphere}
For the single-particle potential inside the Fermi sphere the pertinent integration 
region in the $s\kappa$-plane is the quarter unit disc $s^2+\kappa^2<1$. Taking into 
account this constraint on the variables $(s,\kappa)$ the functions  $F(s,\kappa)$ 
and $ \widehat F(s,\kappa,x)$ introduced in eq.(37) are given by the expressions:   
\begin{equation} F(s,\kappa) = 1+s -\kappa \ln{1+s +\kappa \over 1+s 
-\kappa}+{1\over 2s}(1-s^2-\kappa^2)\ln{(1+s)^2 -\kappa^2 \over 1-s^2 -
\kappa^2}\,, \end{equation}
\begin{eqnarray} \widehat F(s,\kappa,x) &=& {1\over s x}\Bigg\{\theta(2s-1)\,
\theta(2s-1-x)  \ln{|(s+x)^2-\kappa^2|\over |(s-x)^2-\kappa^2|} \nonumber \\ && 
\qquad + \,\theta(x-|2s-1|) \ln{2|(s+x)^2-\kappa^2|\over |1+x^2-2(s^2+\kappa^2)|}
\Bigg\}\,, \end{eqnarray}
\begin{equation}\widehat F(s,\kappa,0)= {4 \,\theta(2s-1)\over s^2-\kappa^2}\,. 
\end{equation}
where $x$ lies in the interval $0<x<1$. Note that a factor $1-\theta(k_f-|2\vec P 
-\vec p\,|)$ is involved in the angular averaging procedure that generates the 
function $ \widehat F(s,\kappa,x)$. This feature requires to study separately 
the cases $0<s<1/2$ and $1/2<s<1$. An interesting relation between these functions 
is: $\int_0^1 dx\,x^2 \widehat F(s,\kappa,x) = F(s,\kappa)-{\rm Re}\, F(-s,\kappa)$.
It involves the difference of two terms, while the corresponding sum is given by: 
$F(s,\kappa)+{\rm Re}\, F(-s,\kappa)= R(s,\kappa)$.

Following the construction of the complex single-particle potential by means of the 
interaction density as described in section 2.2, one obtains from $V_p(\tilde \eta)$ 
in eq.(39) the following double-integral representations for the real and imaginary 
part:
\begin{equation} U(p,k_f)= {8 a k_f^3\over M}\int\limits_0^1 \!ds\, s^2  
\!\!\int\limits_0^{\sqrt{1-s^2}}  \!\!d\kappa \, \kappa  \,\Bigg\{
{a k_f \widehat F(s,\kappa,x)I(s,\kappa)\over [\pi+ a k_f F(s,\kappa)]^2} 
-{\widehat I_*(s,\kappa, x)\over \pi+ ak_f F(s,\kappa)} \Bigg\} \,,\end{equation}
\begin{equation} W(p,k_f)= {8\pi a^2 k_f^4\over M}\int\limits_0^1 \!ds\, s^2  
\!\!\int\limits_0^{\sqrt{1-s^2}}  \!\!d\kappa \, \kappa  \,{[\widehat I_*(s,\kappa,x)
-\widehat I(s,\kappa,x)]I(s,\kappa)\over [\pi+ a k_f F(s,\kappa)]^2} \,.
\end{equation}
The denominator functions in eqs.(43,44) possess lines of zeros 
and therefore they have to be interpreted as (regularized) distributions: 
$X^{-\nu} = {\rm Re}\lim_{\epsilon\to 0} (X+ i \epsilon)^{-\nu}$ for $\nu =1,2$. The 
line of zeros of the function $F(s,\kappa)$ inside the unit quarter disc is 
shown be the (upper) dashed line in Fig.\,3. This curve starts at $s_0=0$, 
$\kappa_0= 0.83356$ and ends at $s_1=0.40865$, $\kappa_1= 0.91269$. 

In order to prove the validity of the Hugenholtz-Van-Hove theorem for the 
single-particle potential $U(k_f,k_f)$ at the Fermi surface, the following 
identity is now instrumental:
\begin{equation} {\partial \over \partial k_f} \Big[ k_f F(P/k_f,q/k_f) \Big]= 
{1\over s}\, \ln{(s+1)^2-\kappa^2\over 1-s^2-\kappa^2} = \widehat F(s,\kappa,1) 
\,. \end{equation}
\begin{figure}[h]
\begin{center}
\includegraphics[scale=0.4,clip]{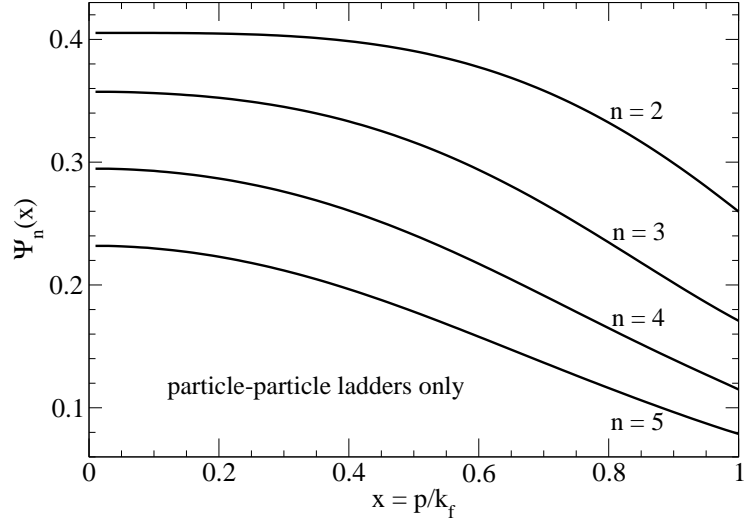}
\end{center}
\vspace{-.8cm}
\caption{Momentum dependence of perturbative contributions to the real 
single-particle potential inside the Fermi sphere.}
\end{figure}

\begin{figure}[h]
\begin{center}
\includegraphics[scale=0.4,clip]{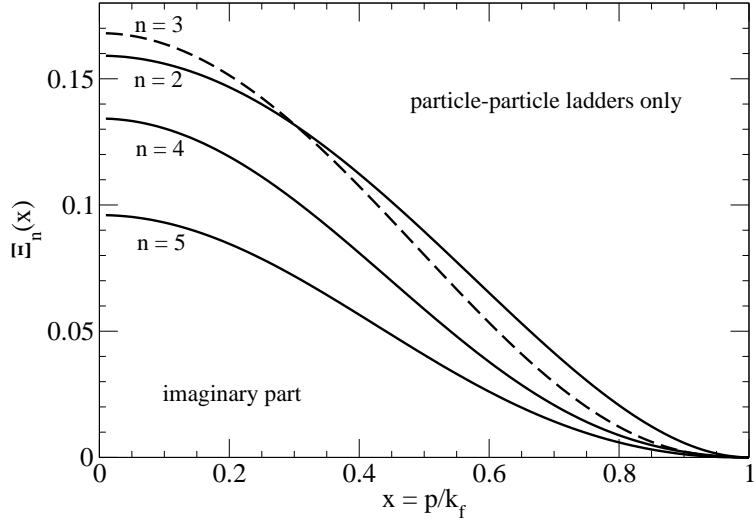}
\end{center}
\vspace{-.8cm}
\caption{Momentum dependence of perturbative contributions to the imaginary 
single-particle potential inside the Fermi sphere.}
\end{figure}
The perturbative expansion of the complex single-particle potential 
$U(p,k_f)+i\,W(p,k_f)$ given in eqs.(43,44) has again the form:  
\begin{equation} U(p,k_f)+i\,W(p,k_f)= {k_f^2 \over 2M} \sum_{n=1}^\infty (-a k_f)^n  
\Big[\Psi_n(x)+i\, \Xi_n(x)\Big]\,,  \end{equation} 
where e.g. the dimensionless functions $\Psi_n(x)$ for the real part can be 
calculated as:
\begin{equation}\Psi_n(x) = {16 \over \pi^n} \int\limits_0^1 \!ds\, s^2  
\!\!\int\limits_0^{\sqrt{1-s^2}}  \!\!d\kappa \, \kappa  \,F(s,\kappa)^{n-2} \Big[ 
(n-1) \widehat F(s,\kappa,x)I(s,\kappa)+  \widehat I_*(s,\kappa, x) F(s,\kappa)
\Big]\,.\end{equation} 
The functions at the lowest two orders are: $\Psi_1(x) = 4/3\pi$, 
$\Xi_1(x)=0$ and $\Psi_2(x)=\Phi_2(x)$,   $\Xi_2(x)=\Omega_2(x)$, with the analytical 
expressions for $\Phi_2(x)$ and $\Omega_2(x)$ written in eqs.(28,30) for $0<x<1$. 
These equalities come from the fact that hole-hole ladder diagrams \cite{schaefer} start 
to contribute to the energy density\footnote{In the conventional counting the 
single-particle potential at second order includes also hole-hole 
contributions.} first 
at order $a^3$. One verifies that the 
double-integral representation of $\Psi_2(x)$ in eq.(47) gives results that are 
in perfect agreement with the analytical expression for $\Phi_2(x)$. Note that in 
the present calculation the identity $\Phi_2(x)=\Psi_2(x)$ is not self-evident, 
since different integrands $\widehat R(s,\kappa,x) I(s,\kappa)+ \widehat 
I_*(s,\kappa,x) R(s,\kappa) \ne  \widehat F(s,\kappa,x) I(s,\kappa)+ \widehat 
I_*(s,\kappa,x) F(s,\kappa)$ are used to represent both functions. 

\begin{figure}[h]
\begin{center}
\includegraphics[scale=0.4,clip]{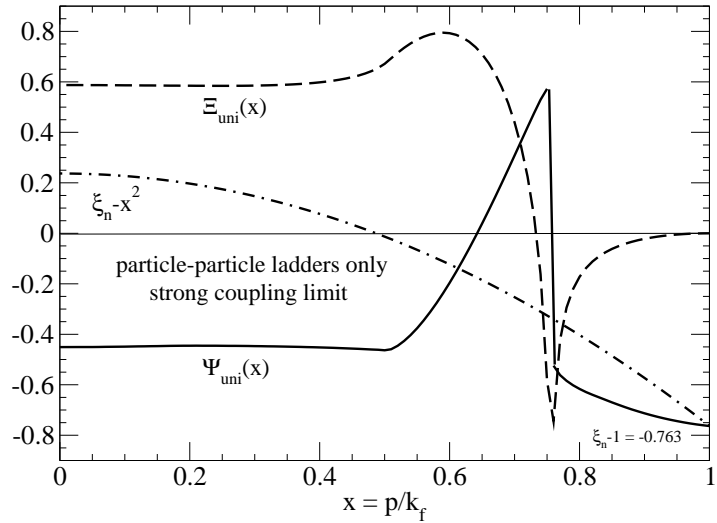}
\end{center}
\vspace{-.8cm}
\caption{Momentum dependence of the single-particle potential inside the Fermi sphere
in the limit $a\to \infty$.}
\end{figure}

The $x$-dependence of the functions $\Psi_n(x)$ and  $\Xi_n(x)$ is shown for 
$n=2,3,4,5$ in Figs.\,10,11. One observes monotonically decreasing functions of $x$ 
as well as a tendency that higher-orders get smaller in magnitude. Precise 
numerical values of the boundary values $\Psi_n(0)$, $\Xi_n(0)$ and $\Psi_n(1)$ are 
listed in Table\,1 up to order $n=8$. The right boundary values are determined by 
the Hugenholtz-Van-Hove theorem as: $\Psi_n(1)= 16(n+5)\pi^{-n} 
\int\!ds\,s^2\!\int\!d\kappa\,\kappa\,I(s,\kappa)F(s,\kappa)^{n-1}$. Note that for 
$n\geq 3$ the functions $\Phi_n(x)$ (shown in Fig.\,4 are always larger than the 
functions $\Psi_n(x)$. At each order their difference is a measure of the additional 
contributions from the (combined) hole-hole ladder diagrams.

It is again straightforward to perform the limit $a \to -\infty$ of 
the single-particle potentials $U(p,k_f)$ and $W(p,k_f)$ given in eqs.(43,44). 
The corresponding results read:
\begin{equation} U(p,k_f)^{(\infty)} = {8 k_f^2\over M}\int\limits_0^1 \!ds\, 
s^2  \!\!\int\limits_0^{\sqrt{1-s^2}}  \!\!d\kappa \, \kappa  \,\Bigg\{
{\widehat F(s,\kappa,x)I(s,\kappa)\over F(s,\kappa)^2}-{\widehat I_*(s,\kappa, x) 
\over F(s,\kappa)} \Bigg\} = {k_f^2 \over 2M} \,\Psi_{\rm uni}(x) \,,\end{equation}
\begin{equation} W(p,k_f)^{(\infty)} = {8 \pi k_f^2\over M}\int\limits_0^1 \!ds\, 
s^2  \!\!\int\limits_0^{\sqrt{1-s^2}}  \!\!d\kappa \, \kappa  \,{[\widehat 
I_*(s,\kappa,x)-\widehat I(s,\kappa,x)]I(s,\kappa)\over F(s,\kappa)^2}
= {k_f^2 \over 2M}\, \Xi_{\rm uni}(x) \,,\end{equation}
where the functions $\Psi_{\rm uni}(x)$ and $\Xi_{\rm uni}(x)$ describe the dependence 
on the rescaled momentum variable $x=p/k_f$. The calculated boundary values  are:
\begin{equation}\Psi_{\rm uni}(0)=-0.451 \,, \qquad \Psi_{\rm uni}(1) = \xi_n^{(pp)} 
-1 = -0.763\,, \qquad \Xi_{\rm uni}(0) = 0.587\,, \end{equation}
with $\xi_n^{(pp)} = 0.237$ the normal Bertsch parameter \cite{resum1,schaefer} obtained 
from the resummed particle-particle ladder diagrams.\footnote{For the resummed 
particle-particle ladders the parameter $\zeta$ (defined in sec.\,6 of 
ref.\cite{resum1}) has the value $\zeta = 1.489$.} The 
$x$-dependence of the function $\Psi_{\rm uni}(x)$ is shown in Fig.\,12. In the 
region $0.5 <x<0.8$ one finds an extreme behavior. After a steep rise up to 
positive values of $0.58$ follows a discontinuous drop-back to negative values at 
$x =0.75$. It is a challenge to evaluate with good numerical accuracy the 
integrals over the regularized double-pole in eq.(48). However, the employed 
methods\,\footnote{All integrals have been 
computed numerically with Mathematica using the method ``LocalAdaptive''.} showed 
good convergence in the range $\epsilon = 10^{-2}\dots 10^{-3}$ of the regulator 
parameter and could be successfully tested with analytically solvable 
examples. The prompt reproduction of the relation $\Psi_{\rm uni}(1) = \xi_n^{(pp)} -1$  
as imposed by the Hugenholtz-Van-Hove theorem is a further test of the quality of the 
numerical methods. The negative slope of $\Psi_{\rm uni}(x)$ at $x=1$ translates 
into a slightly enhanced effective mass: $M^*/M=[1+\Psi'_{\rm uni}(1)/2]^{-1}\simeq 
1.2$. The average value of $\Psi_{\rm uni}(x)$ taken over the interior of the Fermi 
sphere is: $3\int_0^1 dx\,x^2 \Psi_{\rm uni}(x) = -48 \int ds\,s^2 \int d\kappa \,
\kappa\, I_* R/F^2= -0.433$. This is rather close to the value $\Psi_{\rm uni}(0) = 
-0.451$ at the bottom of the Fermi sea.
The dashed line in Fig.\,12 shows the $x$-dependence of the function 
$\Xi_{\rm uni}(x)$ associated with the imaginary potential. While having positive 
values over a wide range in $x$, it changes its sign in the vicinity of the point 
where $\Psi_{\rm uni}(x)$ has a discontinuity. In the calculation the negative values 
of $\Xi_{\rm uni}(x)$ result from treatment of the double-pole $1/F(s,\kappa)^2$ 
in eq.(49) as a regularized distribution. It should be stressed that the same 
regularization in the case of the real part $U(p,k_f)^{(\infty)}$ reproduces the 
relation $\Psi_{\rm uni}(1)=\xi_n^{(pp)}-1$ as imposed by the Hugenholtz-Van-Hove theorem.
The negative values of $\Xi_{\rm uni}(x)$ for $0.74<x<1$ imply an instability of the 
system against excitations of hole states in the Fermi sea. This instability should 
be viewed as a failure of the truncation to particle-particle ladder diagrams only. 
Moreover, there exists the momentum region $0.61<x<0.76$ inside which 
$\Psi_{\rm uni}(x)>\xi_n^{(pp)}-x^2$ holds and an instability to bubble formation 
in the Fermi sea occurs.
\subsection{Continuation into region outside the Fermi sphere}
In this subsection we present the modifications of the formalism that are
necessary in order to continue the complex single-particle potential $U(p,k_f)+i\,
W(p,k_f)$ generated by the particle-particle ladder diagrams into the region $p>k_f$ 
outside the Fermi sphere. 

\begin{figure}[h]
\begin{center}
\includegraphics[scale=0.4,clip]{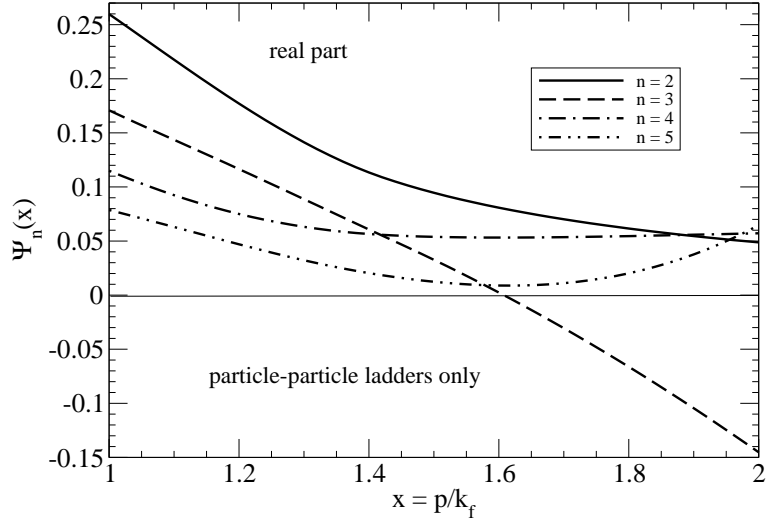}
\end{center}
\vspace{-.8cm}
\caption{Momentum dependence of perturbative contributions to the real 
single-particle potential outside the Fermi sphere.}
\end{figure}

\begin{figure}[h]
\begin{center}
\includegraphics[scale=0.4,clip]{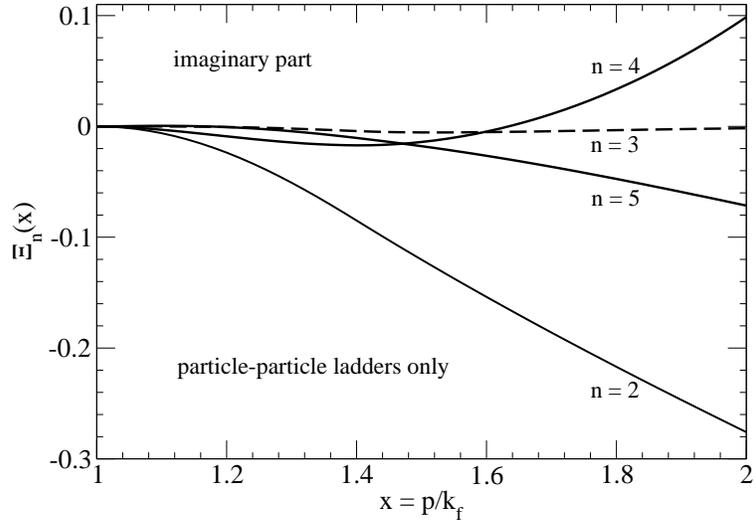}
\end{center}
\vspace{-.8cm}
\caption{Momentum dependence of perturbative contributions to the imaginary 
single-particle potential outside the Fermi sphere.}
\end{figure}

Viewed as a weighting function for integrals over $\theta(k_f-|\vec p_1|)\,\delta^3
(\vec p_2 -\vec p\,)$, the support of the function $\widehat I_*(s,\kappa,x)$  
consists for $x>1$ only of a circular wedge (defined by the inequalities 
$s+\kappa>x$ and  $s^2+\kappa^2<(x^2+1)/2$). Consequently, this function takes for 
$x>1$ the simpler form:
\begin{equation} \widehat I_*(s,\kappa,x) = {1\over s x}\, \theta(s+\kappa-x)
 \, \theta(1+x^2-2(s^2+\kappa^2))\,. \end{equation}
The largest possible values of $s$ and $\kappa$ are now $s_+=\kappa_+
= (x+1)/2>1$. For values $s>1$ the two Fermi spheres, which define the integration 
region of the in-medium loop $B_p$ in eq.(36), are no longer overlapping but get 
completely separated. This implies a change in the lower integration 
boundaries of both the radial and angular coordinates of $\vec l$. Taking into 
account the modifications which occur for $s>1$, the extended function 
$F(s,\kappa)$ reads: 
 \begin{eqnarray} F(s,\kappa) &=& 1+s -\kappa \ln{1+s +\kappa \over 1+s 
-\kappa}+{1\over 2s}(1-s^2-\kappa^2)\ln{(1+s)^2 -\kappa^2 \over |s^2 +
\kappa^2-1|}  \nonumber \\ && +\, \theta(s-1)\,\Bigg\{1-s+\kappa\ln{|s +\kappa-1|
 \over |s -\kappa-1|}+{1\over 2s}(s^2+\kappa^2-1)\ln{|(s-1)^2 -\kappa^2| 
\over |s^2 +\kappa^2-1|} \Bigg\}\,. \end{eqnarray}
Actually, for $s>1$ the equality $F(s,\kappa)= R(s,\kappa)$ holds (see eq.(6)). 
The other function $\widehat F(s,\kappa,x)$ changes also when $x>1$, and its 
continued version reads: 
\begin{equation} \widehat F(s,\kappa,x) = {1\over s x}\Bigg\{\theta(
x-2s-1)\ln{|(s+x)^2-\kappa^2|\over |(s-x)^2-\kappa^2|}  + \theta(2s+1-x) 
\ln{2|(s+x)^2-\kappa^2|\over |1+x^2-2(s^2+\kappa^2)|} \Bigg\}\,. \end{equation}
Note that the distinction of cases $0<s<1/2$ and $1/2<s<1$ is no longer necessary, 
since the case $x<1-2s$ becomes inapplicable when $x>1$. Returning to the resummed
interaction density $V_p(\tilde \eta)$ in eq.(39) and using the two extended 
functions $F(s,\kappa)$ and $\widehat F(s,\kappa,x)$,  the continuation of the 
complex single-particle potential $U(p,k_f)+i\,W(p,k_f)$ into the region $p>k_f$ 
outside the Fermi sphere takes the form: 
\begin{eqnarray} U(p,k_f)&=& {8 a k_f^3\over M}\int\limits_0^{(x+1)/2} \!ds\, s^2  
\!\!\int\limits_0^{(x+1)/2}  \!\!d\kappa \, \kappa  \,\Bigg\{
{a k_f \widehat F(s,\kappa,x)I_*(s,\kappa)\over [\pi+ a k_f F(s,\kappa)]^2} 
-{\widehat I_*(s,\kappa, x)\,\theta(1-s^2-\kappa^2) \over \pi+ ak_f F(s,\kappa)} 
\nonumber \\ && \qquad  \qquad \qquad \qquad \qquad  \qquad  
- \,{\widehat I_*(s,\kappa, x)[\pi+ ak_f F(s,\kappa)]\,\theta(s^2+\kappa^2-1) \over 
[\pi+ ak_f F(s,\kappa)]^2+[ak_f\pi\, I(s,\kappa)]^2} \Bigg\} \,,\end{eqnarray}
\begin{equation} W(p,k_f)= -{8\pi a^2 k_f^4\over M}\int\limits_0^{(x+1)/2} \!ds\, s^2  
\!\!\int\limits_0^{(x+1)/2}  \!\!d\kappa \, \kappa  \,{\widehat I_*(s,\kappa, x)
I(s,\kappa)\,\theta(s^2+\kappa^2-1) \over [\pi+ ak_f F(s,\kappa)]^2+[ak_f\pi\, 
I(s,\kappa)]^2}\,. \end{equation}
Note that the imaginary part Im\,$\bar B_p= ak_f[ I(s,\kappa)-I_*(s,\kappa)]$ at  
$\tilde \eta =0$, which is nonzero only for $s^2+\kappa^2>1$, affects both the 
real and the imaginary single-particle potential outside the Fermi sphere. The last 
term in eqs.(54,55) indicates that the complex denominator $\pi+a k_f[F(s,\kappa)
- i\pi\,I(s,\kappa)]$ has been involved in the derivation.

We first analyze the continuation of $U(p,k_f)+i\,W(p,k_f)$ into the region $p>k_f$ 
according to the perturbative expansion in eq.(46). For the contribution at second 
order one has again the equalities: $\Psi_2(x)=\Phi_2(x)$ and $\Xi_2(x)=\Omega_2(x)$, 
with the corresponding analytical expressions for $x>1$ given in eqs.(29,30). The 
double-integral representation of the function $\Psi_2(x)$ for $x>1$ as obtained by 
expanding eq.(54) to second order in $ak_f$ leads to results which are in perfect 
numerical agreement with $\Phi_2(x)$ written in eq.(29). This consistency serves as 
an important check on the formalism (i.e. modified functions $F(s,\kappa)$ and 
$\widehat F(s,\kappa,x)$) employed in the continuation of the complex single-particle 
potential $U(p,k_f)+i\,W(p,k_f)$ into the region $p>k_f$. The $x$-dependence of the 
functions $\Psi_n(x)$ and $\Xi_n(x)$ with $n=2,3,4,5$ is shown in Figs.\,13,14 for 
$1<x<2$. One observes a behavior quite similar to that of the functions $\Phi_n(x)$ 
and $\Omega_n(x)$ shown in Figs.\,6,7. This means that the additional (combined) 
hole-hole ladder diagrams have little influence on the properties of particle 
excitations outside the Fermi sphere.  

Finally, the continuation of the functions $\Psi_{\rm uni}(x)$ and $\Xi_{\rm uni}(x)$ 
describing the momentum dependence of the complex single-particle potential 
in the limit $a\to \infty$ is shown in Fig.\,15 for $1<x<3$. Again, some 
non-trivial numerics is involved in producing the curve for $\Psi_{\rm uni}(x)$ in 
Fig.\,15. One observes a fast decrease of the attractive real part in the region 
$1<x< 2$. Particle excitations with higher momentum $x> 2.1$ experience a weak 
repulsion from the fermionic medium. The behavior of the function 
$\Xi_{\rm uni}(x)$ for $x>1$ is very similar to that of $\Omega_{\rm uni}(x)$ shown in 
Fig.\,9. In fact for $x\geq 3$ both functions become equal: $\Xi_{\rm uni}(x)= 
\Omega_{\rm uni}(x)$. The reason for this equality is that for $x>3$ only values of 
$s>(x-1)/2>1$ contribute to the double-integral. Under this condition one has 
$F(s,\kappa) = R(s,\kappa)$ and the representations of $W(p,k_f)$ in eqs.(21,55) 
coincide with each other. It should be noted that the negative values of  
$\Xi_{\rm uni}(x)$ for $x>1$ ensure the stability of particle excitations outside 
the Fermi sphere. When approaching the Fermi surface from the right, the (negative) slope 
$\Xi'_{\rm uni}(1^{\!+})= -8\pi(1-s_1)s_1^2 = -2.4819$ (with $s_1=0.40865$) is 
found. At finite coupling strength $ak_f$, one obtains for the left and right 
slope of the function $\Xi(x) = 2M k_f^{-2}\, W(p,k_f)$ at the Fermi surface $x=1$:
\begin{equation} \Xi'(1^{\!-})=0\,, \qquad  \Xi'(1^{\!+})=- {8\pi s_1^2(1-s_1^2)
\over 1+s_1+\pi(ak_f)^{-1}s_1^2}\,,\end{equation}   
where $s_1$ is the solution of the equation $F(s_1,\sqrt{1-s_1^2}) =-\pi/ak_f$.
The largest negative slope  $\Xi'(1^{\!+})\simeq - 8\pi/3$ is reached for 
$ak_f \simeq -\pi/2$ (approached from below).

\begin{figure}[h]
\begin{center}
\includegraphics[scale=0.4,clip]{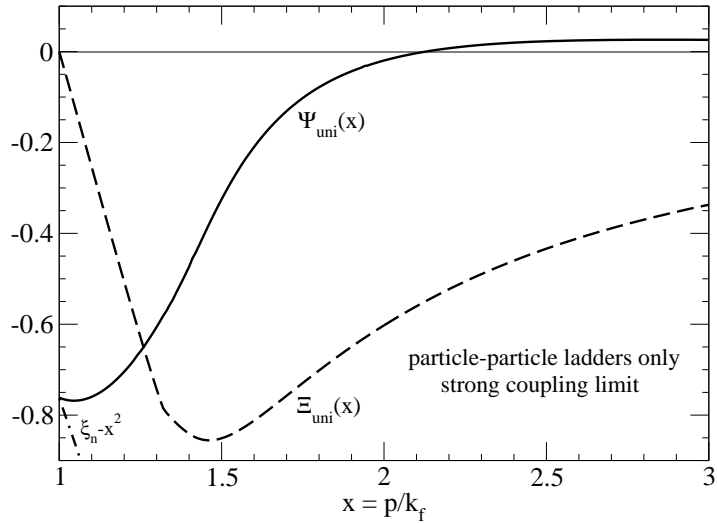}
\end{center}
\vspace{-.8cm}
\caption{Momentum dependence of the single-particle 
potential outside the Fermi sphere in the limit $a\to \infty$. }
\end{figure}

Furthermore, we have extended the construction of the complex single-particle potential   
$U(p,k_f)+i\,W(p,k_f)$ to the hole-hole ladder diagrams \cite{schaefer}. At leading 
order $a^3$, one obtains precisely the difference between the third order results in 
section\,2 and section\,3, proportional to $(-ak_f)^3$. Beyond that order mixed ladder
diagrams with so far unknown systematics appear.

\section{Single-particle potential from ring diagrams}
In this section we calculate the on-shell single-particle potential which arises from 
the particle-hole ring diagrams generated by a contact-interaction proportional 
to the scattering length $a$. We follow the method to treat ring diagrams as 
described in the text book by Gross, Runge and Heinonen \cite{gross}. The key 
quantity is the (euclidean) polarization function:  
\begin{equation} B_{\rm ring} = 8\pi a \int\!\!{d^3 l \over (2\pi)^3} \, {2\vec l 
\cdot \vec q \over (\vec l \cdot \vec q\, )^2 + M^2 \omega^2} \,
N_\eta(\vec l -\vec q/2, \vec p\,) \Big[1- N_\eta(\vec l +\vec q/2, \vec p\,)
\Big]\,, \end{equation}
where the integrand is the Fourier transform of the exponential term in 
eq.(22.15) of ref.\cite{gross} and the factor $N_\eta(\dots)[1- N_\eta(\dots)]$ 
incorporates the pertinent integration boundaries together with the perturbation 
by the test-particle of momentum $\vec p$. Averaging over the directions of 
$\vec p$, one gets to linear order in the parameter $\tilde\eta=\eta\,\pi^2/k_f^3$: 
\begin{equation} \bar B_{\rm ring}= {2 ak_f\over \pi} \Big\{ Q(s,\kappa)+ 
\tilde\eta \, \widehat Q(s,\kappa,x)\Big\}\, \,, \end{equation}
where the momenta have been set to $|\vec q\,| = 2s k_f$ and $|\vec p\,| = x k_f$, 
and the frequency to $\omega = 2s\kappa k_f^2/M$. 
The first function introduced in eq.(58) has the well-known form \cite{gross}: 
\begin{equation} Q(s,\kappa) = 1 -\kappa \arctan{1+s\over\kappa}
 -\kappa \arctan{1-s\over\kappa}+ {1\over 4s}(1-s^2+\kappa^2)  
\ln {(1+s)^2+\kappa^2 \over (1-s)^2+\kappa^2} \,,\end{equation}
and its values lie in the range $0<Q(s,\kappa)\leq 2$. The correction term linear 
in $\tilde \eta$ is determined by the second function: 
\begin{equation} \widehat Q(s,\kappa,x) = {1\over 2s x}\ln 
{(s+x)^2+\kappa^2 \over (s-x)^2+\kappa^2} \,, \qquad\qquad 
\widehat Q(s,\kappa,0)= {2\over s^2+\kappa^2}\,. \end{equation} 
An interesting relation between these two functions is: $\int_0^1 dx\,x^2
\widehat Q(s,\kappa,x)= Q(s,\kappa)$. It is worth mentioning that exactly the 
same result for $\bar B_{\rm ring}$ is obtained, if the last factor $1-N_\eta(\dots)$ 
in eq.(57) is dropped. Such a simpler representation of the polarization function
$B_{\rm ring}$ can be derived in the finite temperature formalism when taking the 
limit $T\to 0$ in the end.    

The interaction energy per particle arising from the leading $n$-ring diagram 
(with a spin-factor $2^n$) is given by the expression:
\begin{equation} \bar E(k_f)^{n-\rm ring} = -{12 k_f^2 \over \pi M n} 
\bigg({2a k_f\over \pi}\bigg)^{\!n} \int\limits_0^\infty \! ds\,s^3 
\int\limits_0^\infty\! d\kappa\, \big[Q(s,\kappa)\big]^n = 
{k_f^2 \over 2M}(ak_f)^n \, \Gamma_n\,. \end{equation}
All the additional exchange-type diagrams due to the Pauli exclusion-principle 
can be taken into account simply by replacing the factor $2^n$ by the proper 
spin-factor $S_n=1+3(-1)^n$ \cite{schaefer}. Starting at third order and going 
up to 8-th order, the numerical values of the coefficients $\Gamma_n$ are: $\Gamma_3 
= -0.22955$, $\Gamma_4 = -0.069116$, $\Gamma_5 = -0.038274$, $\Gamma_6 = -0.025904$, 
$\Gamma_7 = -0.019548$, $\Gamma_8 = -0.015841$. Note that these coefficients are 
all negative and they decrease in magnitude.

\begin{figure}[ht]
\begin{center}
\includegraphics[scale=0.4,clip]{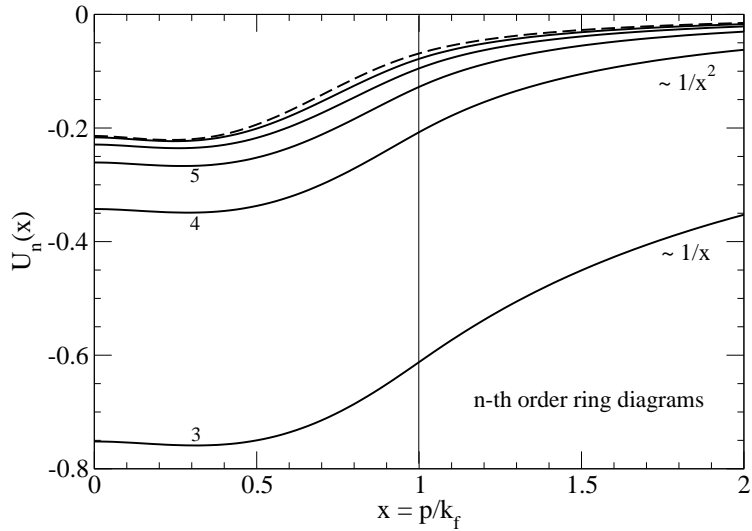}
\end{center}
\vspace{-.8cm}
\caption{Momentum dependence of contributions from particle-hole ring diagrams
to the real single-particle potential.}
\end{figure}

Since the interaction density of a $n$-ring diagram is completely determined by the 
$n$-th power of the polarization bubble $\bar B_{\rm ring}$ the expression for the 
on-shell single-particle potential follows as:
\begin{equation} U(p,k_f)^{n-\rm ring} = -{4k_f^2 \over \pi M } 
\bigg({2a k_f\over \pi}\bigg)^{\!n} \int\limits_0^\infty \! ds\,s^3 
\int\limits_0^\infty\! d\kappa\,\widehat Q(s,\kappa,x) \big[Q(s,\kappa)
\big]^{n-1} = {k_f^2 \over 2M}(ak_f)^n \, U_n(x)\,, \end{equation}
where the dimensionless function $U_n(x)$ describes the momentum dependence 
at $n$-th order.

Fig.\,16 shows the calculated potential-functions $U_n(x)$ in the region $0<x<2$
from third up to 8-th order. One observes an almost constant behavior in the 
region $0<x<0.4$ that is followed by a steady increase and an asymptotic approach to 
zero. The detailed asymptotic behavior of the functions $U_n(x)$ for large $x$ is: 
$U_3(x) \simeq -16(4+\pi^2)/(9\pi^3x)\simeq -0.79523/x$ and  $U_n(x)\simeq 4(n-1) 
\Gamma_{n-1}/(3\pi x^2)$ for $n\geq 4$. The Hugenholtz-Van-Hove theorem is fulfilled 
order by order in the form of the relation $U_n(1) = \Gamma_n(n+5)/3$, which holds 
with very good numerical accuracy. Another interesting property is given by the 
relation $3\int_0^1 dx\,x^2 U_n(x) = n \Gamma_n$. Such a connection between the 
single-particle potential and the contribution to $\bar E(k_f)$ is typical for an 
$n$-body interaction treated at first order in perturbation theory \cite{pionpot}. 
The absence of an imaginary single-particle potential $W(p,k_f)$ from 
particle-hole ring diagrams is also remarkable. We do not study here in further detail 
the resummation of ring diagrams to all orders since for this subclass the limit 
$a \to -\infty$ does not exist.  

\section*{Appendix: Resummation of in-medium ladder diagrams in 
two dimensions}
In this appendix we present the resummation of fermionic in-medium ladder diagrams 
to all orders in two spatial dimensions. The derivation follows essentially the same 
steps as in the case of the three-dimensional calculation presented in detail in 
ref.\cite{resum1}. 
 
For a system of spin-1/2 fermions in two dimensions the relation between particle 
density (per area) and Fermi momentum is $\rho_2 = k_f^2/2\pi$. Likewise, the free 
Fermi gas energy per particle is $\bar E(k_f)^{(0)} = k_f^2/4M$, with $M$ the 
large fermion mass. We introduce a two-body contact-interaction with 
coupling constant $C_0 = 2\pi \alpha/M$, where $\alpha $ is a dimensionless 
parameter. The corresponding first-order Hartree-Fock contribution is readily 
calculated as: $\bar E(k_f)^{(1)} = -\alpha\, k_f^2/4M$. The calculation 
of higher-order contributions from ladder diagrams and their eventual resummation 
to all orders requires the knowledge of the complex-valued in-medium loop 
$ B_0'+ B_1'+ B_2'$ in two dimensions. The rescattering term $ B_0'$ in vacuum 
(with zero medium-insertions) is given by the expression:
\begin{equation} B_0' = 2\pi \alpha \int\!\!{d^2 l \over (2\pi)^2}\,{1\over 
\vec l^{\,2}- \vec q^{\,2}-i \epsilon} = \alpha \bigg( {i \pi \over 2} -
\ln{|\vec q\,| \over \Lambda} \bigg)\,,\end{equation} 
with $\Lambda$ an ultraviolet cutoff. The bound-state pole in the resummed vacuum 
scattering amplitude $\alpha /(1-B_0')$ fixes the value of the cutoff $\Lambda$. 
By solving the equation $ B_0'=1$ one gets $\Lambda = q_b\, e^{1/\alpha}$, where  $q_b$ 
is the binding momentum related to the two-body binding energy by $E_b =-q_b^2/M<0$. 
The contribution $ B_1'$ from the diagrams with one medium-insertion (see Fig.\,1) 
has a real part of the form:
\begin{eqnarray} {\rm Re}\,B_1' &=& -2\pi  \alpha  \int\!\!{d^2 l \over (2\pi)^2}\,
{1\over \vec l^{\,2}- \vec q^{\,2}}\Big[\theta(k_f-|\vec P+\vec l\,|)+ 
\theta(k_f-|\vec P-\vec l\,|)\Big] \nonumber \\ &=& -\alpha\, {\rm Re}\!\int_0^1
\! d\lambda \, {{\rm sign}(\lambda +s^2-\kappa^2) \over \sqrt{[\lambda -
(s+\kappa)^2][\lambda-(s-\kappa)^2]}}=\alpha\,\Big\{\ln\kappa-H(s,\kappa)\Big\}\,, 
\end{eqnarray}
where we have displayed in the second line the intermediate result obtained after 
angular integration. 
Here, $\vec P =(\vec p_1+\vec p_2)/2$ and  $\vec q =(\vec p_1-\vec p_2)/2$ are the 
half-sum and half-difference of two two-component momenta $|\vec p_{1,2}|<k_f$ inside 
a Fermi disc. The dimensionless variables $s = P/k_f$ and $\kappa = q/k_f$ satisfy 
the constraint $s^2+\kappa^2<1$. The radial integral in eq.(64) can be solved in 
terms of a piecewise defined logarithmic function:
\begin{equation} H(s,\kappa) = 2\, \theta(1-s-\kappa) \ln{ 
\sqrt{1-(s+\kappa)^2} +\sqrt{1 -(s-\kappa)^2} \over 2\sqrt{\kappa}} + 
\theta(s+\kappa-1) \ln s \,.\end{equation} 
The imaginary part of the in-medium loop in two dimensions has the form:
\begin{eqnarray} {\rm Im}(B_0'+B_1'+B_2') &=& 2\pi^2  \alpha  \int\!\!{d^2 l \over 
(2\pi)^2}\,\delta(\vec l^{\,2}- \vec q^{\,2})\,\theta(k_f-|\vec P+\vec l\,|) \,
\theta(k_f-|\vec P-\vec l\,|) \nonumber \\
&=&   {B_2' \over 2i} = \alpha\, J(s,\kappa) \,. \end{eqnarray}
where we have already dropped an analogous term proportional to $[1-\theta(\dots)]\,
[1-\theta(\dots)]$, since its phase space vanishes identically by Pauli-blocking and 
energy conservation (see herefore the detailed discussion in section 3 of 
ref.\cite{resum1}). In a geometrical picture the function $4J(s,\kappa)$ measures 
the arc length of a circle inside the intersection of two shifted circular 
discs and it reads:
\begin{equation} J(s,\kappa) = {\pi \over 2}\, \theta(1-s-\kappa) + \theta
(s+\kappa-1) \arcsin{1-s^2-\kappa^2 \over 2s \kappa}  \,.\end{equation} 
It is interesting to note that $ s\kappa J(s,\kappa)$ serves also as a weighting 
function for integrals over the product of two Fermi discs: 
\begin{equation} \int\limits_{|\vec p_{1,2}|<k_f}\!\!\!\!{d^2 p_1 d^2 p_2\over(2\pi)^4} 
\, f(s,\kappa) = {2k_f^4 \over \pi^3}  \int\limits_0^1 \!ds\, s \!\!\int
\limits_0^{\sqrt{1-s^2}}  \!\!d\kappa \, \kappa  \, J(s,\kappa) \, f(s,\kappa)\,,  
\end{equation}
where the integrand depends on $s = |\vec p_1+\vec p_2|/2k_f$ and  
$\kappa = |\vec p_1-\vec p_2|/2k_f$. Following the instructive diagrammatic analysis 
in section 4 of ref.\cite{resum1} one arrives with these ingredients at the 
following result for resummed interaction energy per particle from ladder diagrams:
\begin{equation} \bar E(k_f)= -{8 k_f^2\over \pi M}\int\limits_0^1 \!ds\, s  
\!\!\int\limits_0^{\sqrt{1-s^2}}  \!\!d\kappa \, \kappa  \, \arctan{J(s,\kappa) 
\over  H(s,\kappa)+\ln(k_f/q_b)} \,. \end{equation}
Note that either the parameter $\alpha$ or the cutoff $\Lambda$ have dropped out in the 
above ratio, since according the relation $\alpha^{-1}+ \ln(k_f/\Lambda) = 
\ln(k_f/q_b)$ only the binding momentum 
$q_b$ remains as a physically relevant scale parameter. The dependence of $\bar 
E(k_f)$ on the dimensionless strength parameter $\gamma = \ln(k_f/q_b)$ can be 
analyzed for large values of $\gamma$ in the form of a power series in 
$\gamma^{-1}$, which reads:
\begin{equation} \bar E(k_f)={k_f^2 \over 4M} \bigg\{-\gamma^{-1}+ \bigg({3\over 4}
-\ln 2\bigg) \gamma^{-2}-0.16079\, \gamma^{-3}-0.06009\, \gamma^{-4} -0.13832\, 
\gamma^{-5} + \dots \bigg\}\,. \end{equation}
When setting $\Lambda = k_f$ the expansion in powers of $\gamma^{-1}$ becomes 
equivalent to the weak coupling expansion in powers of $\alpha$. This way one 
recovers the earlier mentioned first-order Hartree-Fock result. Remarkably, the 
coefficient at second order $\gamma^{-2}$ can be calculated exactly with the (small) 
value $3/4-\ln 2 = 0.056853$.

\begin{figure}[ht]
\begin{center}
\includegraphics[scale=0.4,clip]{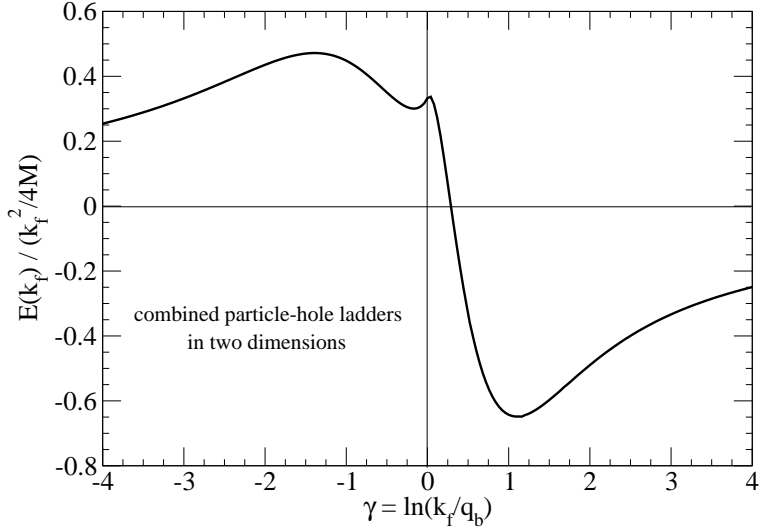}
\end{center}
\vspace{-.5cm}
\caption{Interaction energy per particle $\bar E(k_f)$ divided by the free Fermi 
gas energy $k_f^2/4M$ in two dimensions as a function of the parameter $\gamma = 
\ln(k_f/q_b)$.}
\end{figure}
\begin{figure}[ht]
\begin{center}
\includegraphics[scale=0.4,clip]{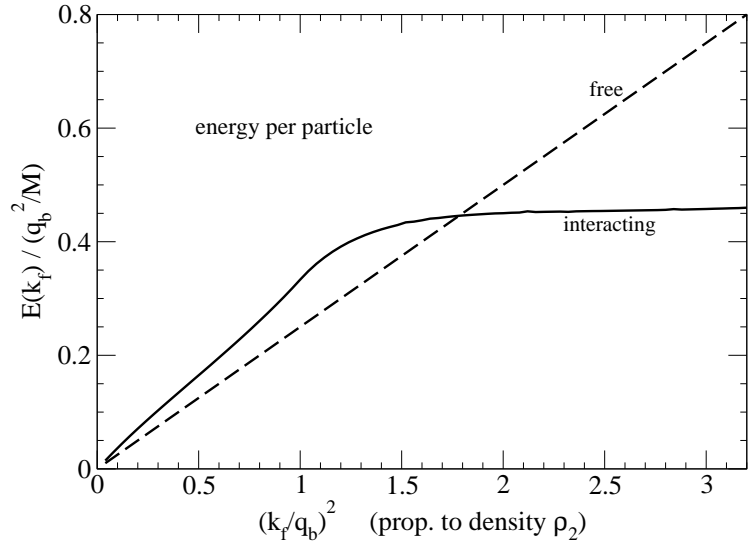}
\end{center}
\vspace{-.5cm}
\caption{Equation of state of a two-dimensional Fermi gas resulting from the 
resummation of ladder diagrams to all orders.}
\end{figure}
\begin{figure}[ht]
\begin{center}
\includegraphics[scale=0.4,clip]{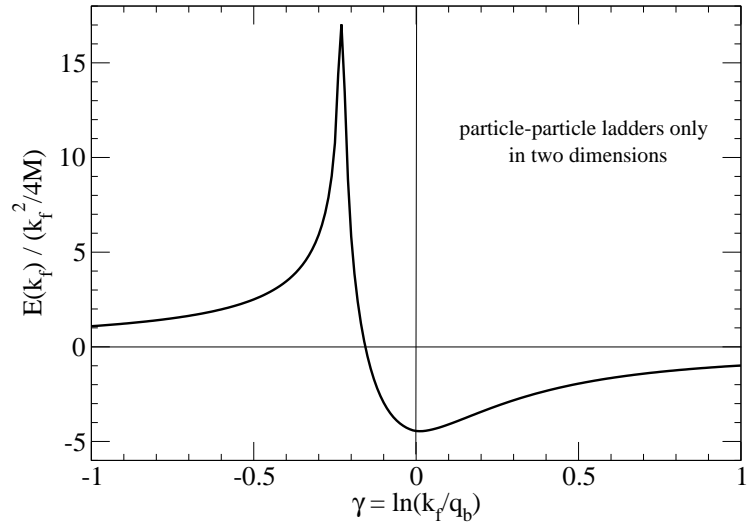}
\end{center}
\vspace{-.5cm}
\caption{Ratio of interaction energy per particle $\bar E(k_f)$ and free Fermi 
gas energy $k_f^2/4M$ as a function of the parameter $\gamma = \ln(k_f/q_b)$. 
Only the particle-particle ladder diagrams are considered.}
\end{figure}
The ratio between the interaction energy per particle $\bar E(k_f)$ and the free 
Fermi gas energy $k_f^2/4M$ is shown in Fig.\,17 as a function of the dimensionless
strength parameter $\gamma = \ln(k_f/q_b)$. One observes an oscillatory behavior with 
moderate repulsion for $\gamma<0.29$ and somewhat stronger attraction 
for $\gamma>0.29$. The asymptotic expansion in eq.(70) provides a very good 
approximation of the ratio $4M \bar E(k_f)/k_f^2$ for the values $|\gamma|>4$ outside 
the plotted range. The positive and negative areas above and below the $\gamma$-axis 
in Fig.\,17 compensate each other exactly. In Fig.\,18 the equation of state of a 
free and an interacting two-dimensional Fermi gas are directly compared with each 
other. The variable $(k_f/q_b)^2$ on the abscissa is now directly proportional to the 
density $\rho_2=k_f^2/2\pi$. One observes weak repulsion in the low-density regime  
and sizeable attraction above a critical density $\rho_c \simeq 0.28 M|E_b|$ that is 
determined by the two-body binding energy .     
  
For the sake of completeness we consider also the particle-particle ladder diagrams 
in two dimensions. The resummation to all orders leads to a geometrical series and  
the corresponding interaction energy per particle is given by the expression: 
\begin{equation} \bar E(k_f)= -{8 k_f^2\over \pi M}\int\limits_0^1 \!ds\, s  
\!\!\int\limits_0^{\sqrt{1-s^2}}  \!\!d\kappa \, \kappa  \, {J(s,\kappa) \over 
 H_p(s,\kappa)+\ln(k_f/q_b)} \,, \end{equation}
with the two-dimensional bubble function: 
\begin{equation} H_p(s,\kappa) = {1\over \pi} \int_0^{\pi/2}\!d\varphi \ln\Big[
\Big(s \cos\varphi+\sqrt{1-s^2\sin^2\varphi}\Big)^2-\kappa^2\Big]\,.\end{equation}
Note that the condition $\kappa^2<1-s^2$ ensures that the argument of the logarithm 
is always positive. Although $ H_p(s,\kappa)$ cannot be expressed in terms of 
elementary functions, the following relation holds: $ H_p(s,\kappa) +{\rm Re}\, 
H_p(-s,\kappa)=   H(s,\kappa) +\ln\kappa$, with $H(s,\kappa)$ written in eq.(65).
The weak coupling expansion of $\bar E(k_f)$ in eq.(71) has the form:
\begin{equation} \bar E(k_f)={k_f^2 \over 4M} \bigg\{- \gamma^{-1}+ \bigg({3\over 4}
-\ln 2\bigg) \gamma^{-2} -0.03147\,  \gamma^{-3} -0.004415\,  \gamma^{-4}-0.005598\,  
\gamma^{-5}+ \dots \bigg\}\,, \end{equation}
with $\gamma = \ln(k_f/q_b)$. Obviously, the first two terms agree with eq.(70) and 
the different coefficients of the higher order terms indicate the influence of the 
hole-hole ladder diagrams. Fig.\,19 shows the ratio of the interaction energy per 
particle $\bar E(k_f)$ divided by the free Fermi gas energy $k_f^2/4M$ in the range 
$-1<\gamma<1$. For $|\gamma|>1$ the asymptotic expansion in eq.(73) provides a very 
good approximation of the $\gamma$-dependence. The sharp peak at $\gamma \simeq 
-0.23$ and large magnitude of the ratio in comparison to Fig.\,17 should be 
considered as an artefact of the truncation to particle-particle ladder diagrams. 
The areas above and below the $\gamma$-axis again compensate each other exactly.
It is a generic feature that the truncation to particle-particle ladder diagrams 
develops a much stronger dependence on a control parameter than the complete 
(particle and hole) ladder series. A further example for this fact is the energy per 
particle from a resummed p-wave contact-interaction shown in Fig.\,5 of 
ref.\cite{resum2}. Throughout this work the results obtained from the complete 
(particle and hole) ladder series are the ones with the higher priority.

\section*{Acknowledgement}
I thank T. Hell for discussions and valuable help in numerical calculations with 
Mathematica.  Informative discussions with R. Schmidt and W. Zwerger are gratefully 
acknowledged.

\end{document}